\pdfoutput=1
\documentclass[10pt,twocolumn]{article}
\usepackage[a4paper,margin=0.75in,columnsep=0.25in]{geometry}
\usepackage{amsmath,amssymb,booktabs,graphicx,microtype,enumitem,tabularx}
\usepackage[authoryear,round]{natbib}
\usepackage[colorlinks=true,allcolors=black,urlcolor=blue]{hyperref}
\setlist[itemize]{leftmargin=*,itemsep=3pt}
\newcommand{\load}{\textsc{Load}}
\newcommand{\abstain}{\textsc{Abstain}}
\title{\textbf{Towards Strategy-Level RSI for Skill-Augmented Agents:\\ Learning When to Reuse Skills from Execution Feedback}}
\author{Shuang Guo$^{1}$\thanks{Code and artifacts: \url{https://github.com/kuafuai/SkillApt}; project page: \url{https://gym.codeflying.net/}.}\hspace{1em}
Xiaobo Sui$^{2}$\quad
Fei Jiang$^{2}$\quad
Ruixin Li$^{2}$\\[2pt]
Manxiang Li$^{2}$\quad
Rong Li$^{2}$\quad
Ziyang Ou$^{3}$\\[4pt]
$^{1}$Central China Normal University\qquad
$^{2}$Beijing Kuafu Technology Co., Ltd.\\[2pt]
$^{3}$Monash University}
\date{}
\begin{document}
\maketitle
\begin{abstract}
Long-running agents accumulate reusable Skills, but a Skill that is semantically relevant to a task is not necessarily worth loading in the current state. We study the applicability question that arises once a candidate Skill is known: should it be loaded in the current state? We propose SkillApt, which uses matched WITH/WITHOUT Skill executions on the same task state as persistent evidence, estimates the conditional marginal utility of the Skill, and chooses \load\ or \abstain\ accordingly. The base model, agent architecture, and Skill contents stay fixed; only the external deployment policy changes. We call this constrained setting strategy-level recursive self-improvement (Strategy-Level RSI).

On 20 Skills and 160 held-out states, as paired evidence accumulates, SkillApt's task success rises from 81.9\% under a cold start to 91.3\%, matching a strong zero-shot LLM controller; yet SkillApt activates Skills on only 26.3\% of states, versus 98.8\% for the zero-shot controller. A hard-candidate study shows that non-optimal Skills mostly leave correctness unchanged while raising execution cost, and occasionally cause correctness harm. An ablation shows that a history recording only WITH success makes the policy load almost everywhere, whereas paired evidence substantially improves selectivity.

These results indicate that relevance is not applicability: the main effect of execution evidence is not to make the model stronger but to change how existing Skills are deployed, moving the system from near-always loading to selective reuse.
\end{abstract}
\section{Introduction}
Long-running enterprise and software-engineering agents increasingly accumulate reusable Skills: procedural artifacts that encode task instructions, tool-use routines, debugging procedures, or multi-step workflows. A large body of work improves agents from their own execution experience, through verbal reflection and experiential memory, distilled reasoning memory, induced or searched workflows, searched agent designs, and self-modifying agent programs \citep{shinn2023reflexion,zhao2024expel,ouyang2025reasoningbank,wang2025awm,zhang2025aflow,hu2024adas,zhang2025dgm}. Reusable Skills in particular have become a standard way to externalize capabilities, with work on how to acquire, retrieve, optimize, and evaluate them \citep{wang2023voyager,su2026sra,yang2026skillopt,skillauditfixed2026,skillaudittrajectory2026,sweskillsbench2026}. A different question arises once a Skill exists: when should an agent reuse a capability it already has?

Our central claim is that \textbf{relevance is not applicability}. A Skill can be semantically relevant to a task yet have little or negative marginal execution utility in the current state: the agent may already solve the task without it, and the extra instructions consume context, add cost, and occasionally turn a correct answer into a wrong one. Neither retrieval scores nor successful WITH-Skill executions distinguish these cases. We therefore separate three layers. \emph{Retrieval} proposes Skills that may be relevant; \emph{applicability} decides whether a candidate is worth using now; \emph{execution} carries out the chosen \load\ or \abstain. SkillApt addresses only the second layer: given a candidate, \emph{should this Skill be loaded in the current state?} It is not a retrieval algorithm. Learning when to invoke a reusable procedure has been studied before. SelSkill learns invocation preferences from invoke/skip pairs with dual-granularity preference learning \citep{selskills2026}, Boundary-Aware Skill Memory (BASM; \citealp{lin2026basm}) attaches explicit applicability boundaries to Skill memory, and routing and calibration methods adapt deployment decisions from feedback \citep{genrouter2026,pilot2025,cabs2026,probecal2024}. We ask a narrower question: how accumulated paired execution outcomes revise the activation policy for a fixed candidate Skill under a fixed executor.

We call this setting \emph{strategy-level RSI} and use the term as an operational framing: the base LLM, agent architecture, and Skill contents stay fixed, while a persistent evidence store $\mathcal E_t\rightarrow\mathcal E_{t+1}$ revises an external deployment policy $\pi_t\rightarrow\pi_{t+1}$. What improves is the Skill deployment policy, not the model parameters or the Skill contents. We do not claim unrestricted RSI, autonomous online self-modification, Skill evolution, or model self-training.

SkillApt's default evidence source is matched WITH/WITHOUT executions: the same task state is run with and without the candidate Skill, both outcomes and the derived utility label are stored, and this record is used to decide whether to load the Skill in new states. It learns from these paired outcomes, not from full execution trajectories. Figure~\ref{fig:framework} separates the offline construction of paired evidence used in our experiments from deployment, where only the selected arm runs; our empirical claims concern evidence-conditioned policy revision under this controlled offline protocol.

Our primary experiment covers 20 Skills from the logical-reasoning benchmark LogicBench \citep{parmar2024logicbench} and the medical-calculation benchmark MedCalc-Bench \citep{khandekar2024medcalc} with 160 held-out states. After 12 paired evidence states per Skill, task success rises from 81.9\% under a cold start to 91.3\%, matching a strong zero-shot LLM controller. The difference lies in deployment: SkillApt activates the Skill on 26.3\% of states, whereas the zero-shot controller does so on 98.8\%.

Our contributions are threefold:
\begin{itemize}
\item We pose a problem distinct from Skill retrieval: once a candidate Skill is known, decide whether it is worth activating in the current state.
\item We propose SkillApt, which uses matched WITH/WITHOUT execution outcomes to estimate the conditional marginal utility of a Skill and chooses \load\ or \abstain\ accordingly.
\item We show that SkillApt matches the task success of a strong zero-shot controller at a much lower Skill activation rate, and that in ablations paired evidence is what keeps Skill exposure selective.
\end{itemize}

\section{Related Work}
\subsection{Experience-Driven Agent Improvement}
Agents have been improved from their own execution experience at several levels \citep{shinn2023reflexion,zhao2024expel,ouyang2025reasoningbank,wang2025awm,zhang2025aflow,hu2024adas,zhang2025dgm}. At the memory level, Reflexion retains verbal feedback, ExpeL extracts reusable insights, and ReasoningBank distills successful and failed trajectories into reasoning memory, none of them changing the underlying model \citep{shinn2023reflexion,zhao2024expel,ouyang2025reasoningbank}. At the workflow level, Agent Workflow Memory induces reusable routines and AFlow searches over workflow structure \citep{wang2025awm,zhang2025aflow}. At the architecture and code level, Automated Design of Agentic Systems (ADAS; \citealp{hu2024adas}) searches agent designs and the Darwin G\"odel Machine \citep{zhang2025dgm} modifies agent programs. Experience can thus act on several different parts of an agent. Strategy-level RSI adds one more, the deployment policy, and restricts what changes: a persistent evidence store revises an external deployment policy, not the model, the procedure, or the agent code.

\subsection{Skill Acquisition, Retrieval, and Evolution}
Reusable procedural artifacts are now a standard way to externalize agent capabilities, and prior work covers how to acquire, retrieve, optimize, and evaluate them \citep{wang2023voyager,su2026sra,yang2026skillopt,skillauditfixed2026,skillaudittrajectory2026,sweskillsbench2026}. Voyager accumulates executable Skills in an open-ended environment \citep{wang2023voyager}; Skill Retrieval Augmentation (SRA; \citealp{su2026sra}) separates retrieval from incorporation and execution; SkillOpt optimizes Skill artifacts from rollout feedback \citep{yang2026skillopt}. Two papers that share the name SkillAudit evaluate Skills rather than construct them: \citet{skillauditfixed2026} move evaluation from fixed task suites to Skill-centered assessment, and \citet{skillaudittrajectory2026} audit paired trajectories to evolve Skills without ground truth. Their baseline-comparison perspective is close to our measurement of marginal utility, although we keep the artifacts fixed. Acquiring, retrieving, optimizing, evaluating, and governing the use of Skills are related but distinct interventions that need not improve the same objective; SkillApt concerns only the last.

\subsection{Selective Skill Invocation and Skill Utility}
Whether a retrieved or memorized procedure should actually be used is increasingly treated as a problem in its own right, both by methods for selective invocation and by empirical studies showing that Skills can add overhead or cause failures \citep{selskills2026,lin2026basm,sweskillsbench2026,skillsHarm2026}. SelSkill builds controlled invoke/skip preference pairs from shared trajectory prefixes and learns invocation preferences at the episode and step level \citep{selskills2026}. SkillApt instead adapts a persistent external reuse policy from historical matched outcomes, and its fixed, low-capacity controller lets us examine task correctness and label-based selectivity separately. BASM attaches explicit applicability conditions, risk cues, avoidance rules, and recovery notes to Skill memory \citep{lin2026basm}, whereas SkillApt induces reuse decisions from observed outcomes; our budgeted BASM-inspired comparison evaluates one explicit-boundary implementation, not the full BASM system.

On the empirical side, SWE-Skills-Bench compares WITH and WITHOUT Skill use in software engineering, including execution overhead and mismatched guidance \citep{sweskillsbench2026}, and \citet{skillsHarm2026} study Skill-induced correctness failures, inefficiency, and differential attribution. SkillApt turns this marginal-utility view into a reuse decision driven by stored evidence, and our hard-negative evaluation separates direct correctness harm from added cost at tied correctness.

\subsection{Adaptive Routing and Calibration}
Routing and calibration methods also adapt deployment decisions from feedback \citep{genrouter2026,pilot2025,cabs2026,probecal2024}. GenRouter routes among fixed agentic image-generation workflows using demand profiling, accumulated experience, utility matching, and cost-aware filtering \citep{genrouter2026}; it selects a workflow, whereas SkillApt decides whether a textual Skill is exposed to a fixed executor. PILOT \citep{pilot2025} combines preference priors with contextual-bandit feedback for budget-aware LLM routing, and CABS \citep{cabs2026} combines surrogate rewards with observed feedback in correlation-aware routing; both choose among models rather than deciding whether to expose a procedure. ProbeCal \citep{probecal2024} calibrates a tool-using agent's uncertainty about tool effectiveness from execution feedback, a calibration objective rather than a reuse policy. SkillApt differs in its object of study: historical matched WITH/WITHOUT outcomes for Skill reuse under a fixed executor, indexed by Skill and executing model, with model parameters and Skill contents held fixed.

\section{Method}
\begin{figure*}[t]
\centering\includegraphics[width=\textwidth]{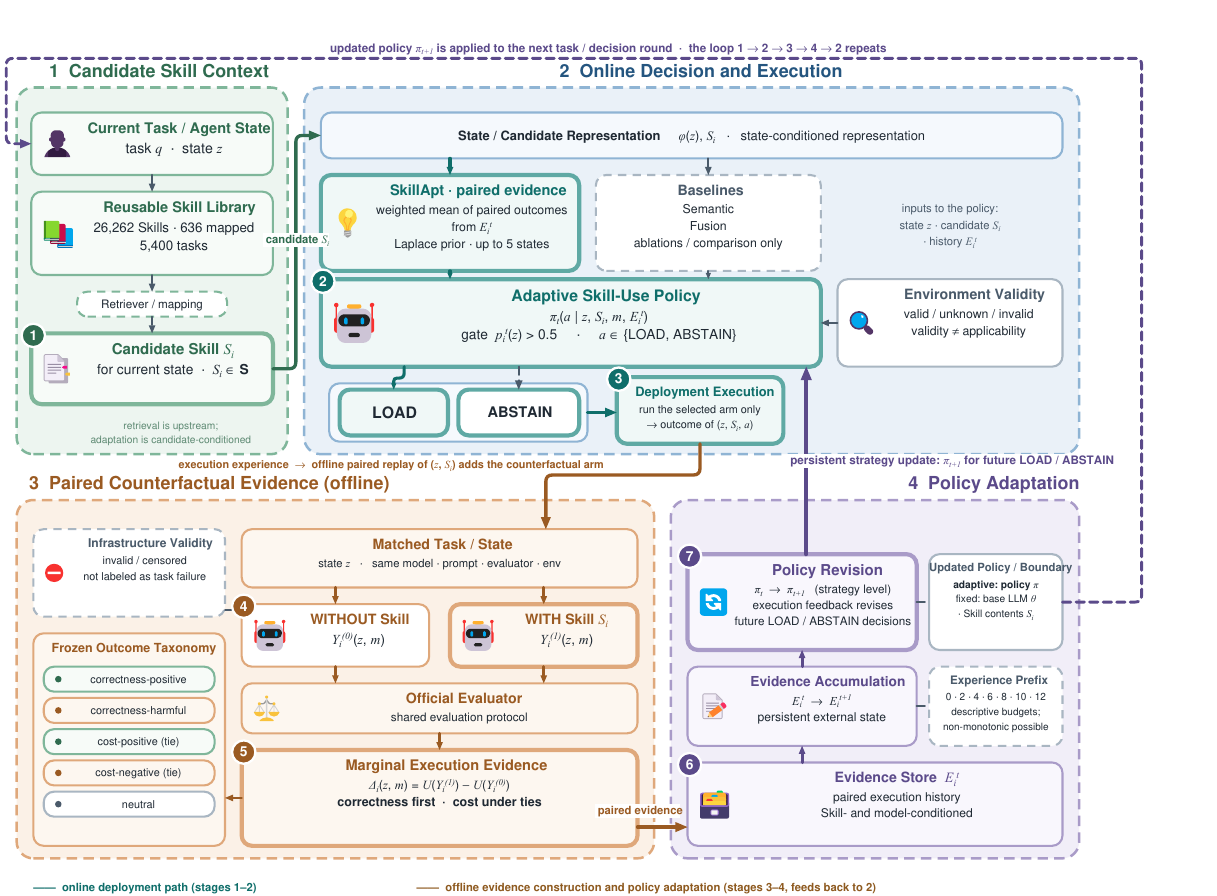}
\caption{SkillApt as a closed-loop strategy-level RSI framework. Stage 1 obtains a candidate Skill $S_i$ for the current state $z$; in Stage 2 the online policy ${\pi_t(a\mid z,S_i,m,\mathcal E_i^t)}$ chooses \load\ or \abstain\ and executes only the selected arm; Stage 3 constructs paired WITH/WITHOUT-Skill execution outcomes for the same $(z,S_i)$ under a fixed model, prompt, evaluator, and environment to estimate the marginal execution utility $\Delta_i(z,m)$; Stage 4 accumulates evidence $\mathcal E_i^t\rightarrow\mathcal E_i^{t+1}$ and updates the policy $\pi_t\rightarrow\pi_{t+1}$ for subsequent tasks. Teal arrows mark the online execution path; orange and violet arrows mark offline evidence construction and policy updating. Model parameters, agent architecture, and Skill contents remain fixed.}\label{fig:framework}
\end{figure*}

\subsection{Strategy-Level RSI Setting}
We use strategy-level RSI as an operational framing: an agent revises a persistent deployment strategy from the consequences of its own earlier executions, while its model parameters and procedural artifacts stay fixed. Concretely,
\begin{equation}
\theta_{t+1}=\theta_t,\quad \mathcal S_{t+1}=\mathcal S_t,\quad
\mathcal E_t\rightarrow\mathcal E_{t+1},\quad\pi_t\rightarrow\pi_{t+1}.
\end{equation}
What self-improves is the external Skill deployment policy, not the model parameters or the Skill contents. The loop is Execution $\rightarrow$ Feedback $\rightarrow$ Persistent Evidence $\rightarrow$ Policy Adaptation $\rightarrow$ Future Execution (Fig.~\ref{fig:framework}). The equations delimit the intervention we study. The experiments evaluate it by offline evidence-prefix replay and a leakage-free sequential replay (Section~\ref{sec:sequential}); they involve no unrestricted RSI, autonomous online self-modification, Skill evolution, or model self-training.

Experience-driven improvement does not have to act on model weights. Language agents have improved without parametric updates by revising verbal reflections \citep{shinn2023reflexion}, extracted insights \citep{zhao2024expel}, or distilled reasoning memory \citep{ouyang2025reasoningbank}; by inducing or searching reusable workflows \citep{wang2025awm,zhang2025aflow}; by searching or rewriting agent designs and code \citep{hu2024adas,zhang2025dgm}; and, in model routing, by updating a persistent selection strategy from observed feedback \citep{pilot2025,cabs2026,genrouter2026}. The Self-Taught Optimizer (STOP; \citealp{zelikman2023stop}) improves a scaffolding program while the underlying language model stays fixed; because the model is unchanged, this is not full recursive self-improvement. Strategy-level RSI applies the same restriction to a narrower object: an external policy that decides whether to expose an existing Skill, conditioned on a persistent store of paired outcomes. In this sense we treat the deployment policy as a target of experience-driven adaptation; none of this prior work studies the applicability decision itself.

\subsection{Problem Formulation}
Let $z$ denote the available task state, $S_i$ a candidate Skill, and $m$ the fixed base model. Retrieval proposes $S_i$ from a library; the applicability decision chooses whether to expose it to the executor; the executor then runs the selected action. Retrieval and applicability are distinct but coupled: candidate quality constrains the available actions, whereas an applicability decision concerns the consequences of using an already proposed candidate. The external policy is
\begin{equation}
\pi_t(a\mid z,S_i,m,\mathcal E_i^t),\quad
a\in\{\load,\abstain\},
\end{equation}
where $\mathcal E_i^t$ is the evidence available for Skill $i$ at stage $t$. Here $t$ indexes accumulated evidence, not necessarily calendar time. In the controlled benchmark, $z$ is represented by the task question. It is not an observed internal model state or a complete software-environment snapshot.

\paragraph{Operational scope of Skills.}We define a Skill broadly as a reusable carrier of procedural capability, which may contain task instructions, tool-use rules, workflow constraints, or executable procedures. The controlled experiments on SRA-Bench, the benchmark introduced with SRA \citep{su2026sra}, operationalize Skills mainly as textual procedural context supplied to the executor through context injection: the WITH condition adds the Skill text to the context but does not force compliance. Executable scripts, code Skills, hard runtime constraints, and tool-plugin Skills are not tested. Our empirical conclusions therefore apply first to textual procedural Skills; script-based, tool-based, and hard-constraint Skills are left to future work.

\subsection{Paired Execution Evidence}
For the same task, model, prompt shell, evaluator, and environment, let $Y_i^{(1)}$ and $Y_i^{(0)}$ denote outcomes with and without the candidate Skill. The conceptual quantity is marginal execution utility,
\begin{equation}
\Delta_i(z,m)=\mathbb E\big[U(Y_i^{(1)})-U(Y_i^{(0)})\mid z,m\big].
\end{equation}
The notation specifies the target comparison; in practice we observe one matched pair per state and apply a frozen correctness-first classification rule, so residual generation variability remains even at temperature zero.

The paired design addresses an identification problem. A successful WITH execution shows that the task was solved while the Skill was present, not that the Skill helped: the same task might have been solved without it. The marginal contribution is the difference between the two outcomes of the same state, the quantity formalized in the potential-outcomes framework \citep{rubin1974estimating}, and any single execution reveals only one of them, which \citet{holland1986statistics} calls the fundamental problem of causal inference. Matched WITH/WITHOUT replay sidesteps this offline by running both arms on the same state under a shared evaluator. Toolformer uses a similar keep-it-only-if-it-helps rule, retaining a candidate API call only when the call and its result lower the language-modeling loss relative to making no call \citep{schick2023toolformer}, and WITH/WITHOUT comparisons are increasingly used to check whether Skills help at all \citep{sweskillsbench2026,skillsHarm2026}. We treat each pair only as a descriptive contrast on one observed sample: arms are not repeated, we do not claim identified causal effects, and the estimator below aggregates noisy contrasts rather than estimating a treatment effect with statistical guarantees.

\paragraph{Why success-only evidence over-attributes.}The identification problem can be made quantitative. For a Skill and a distribution over states, let $\pi_{11},\pi_{10},\pi_{01},\pi_{00}$ denote the probabilities that both arms are correct, that only the WITH arm is correct, that only the WITHOUT arm is correct, and that neither is, with $\pi_{11}+\pi_{10}+\pi_{01}+\pi_{00}=1$. A history that records only WITH success estimates
\begin{equation}
s=\pi_{11}+\pi_{10},\label{eq:withonly}
\end{equation}
whereas the correctness-centered paired target of Eq.~\ref{eq:target} has expectation
\begin{equation}
\bar y=\pi_{10}+\tfrac12(\pi_{11}+\pi_{00})=\tfrac12+\tfrac12(\pi_{10}-\pi_{01}).\label{eq:overattr}
\end{equation}
Hence $\bar y>\tfrac12$ exactly when the Skill's correctness benefit exceeds its correctness harm, $\pi_{10}>\pi_{01}$, whereas $s>\tfrac12$ whenever the task is usually solved with the Skill present, regardless of whether it would have been solved without it. The two statistics differ by $s-\bar y=\tfrac12(\pi_{11}-\pi_{00})$, the mass of states whose outcome does not depend on the Skill. In the Phase A evidence store the four cells are $204$, $23$, $7$, and $6$ of $240$ states, giving $s=0.946$ but $\bar y=0.533$: a success-only history credits the Skill for $204$ of its $227$ recorded successes. This is the attribution bias that the WITH-only ablation in Section~\ref{sec:paired} exhibits empirically. The estimator of Eq.~\ref{eq:score} applies the same contrast locally, with similarity weights and a prior, rather than to the population.

Let $c^{(a)}\in\{0,1\}$ denote evaluator correctness. If $c^{(1)}>c^{(0)}$, the outcome is correctness-positive; if $c^{(1)}<c^{(0)}$, it is correctness-harmful. Let $\mathcal Q=\{\mathrm{tokens},\mathrm{latency},\mathrm{calls}\}$ denote the three resource dimensions. For tied correctness, define the frozen relative cost advantage
\begin{equation}
g_i=\frac{1}{3}\sum_{q\in\mathcal Q}
\frac{C_q^{(0)}-C_q^{(1)}}{\max(1,C_q^{(0)})}.
\label{eq:cost}
\end{equation}
The tied pair is cost-positive if $g_i>0.05$, cost-negative if $g_i<-0.05$, and neutral otherwise. This is an operational multi-resource proxy rather than a monetary cost model: tool-call counts are zero in the direct-generation benchmark but still occupy a term, and latency reflects provider and scheduling variation. We retain the rule for consistency and report tokens separately; a token-only relabeling is given in Appendix~\ref{app:token-only}.

A successful WITH arm therefore does not establish positive marginal utility. When both arms are correct, a longer or slower assisted execution may instead receive a cost-negative label. Similarly, two incorrect arms can differ in cost without either being a successful solution. Infrastructure-invalid arms are excluded from utility labeling rather than treated as incorrect executions.

\subsection{Evidence Accumulation}
SkillApt uses matched paired execution evidence as its default evidence source. Each valid evidence item associates a state with paired outcomes and an estimator target:
\begin{equation}
\mathcal E_i^t=\{(z_{ij},Y_{ij}^{(0)},Y_{ij}^{(1)},y_{ij})\}_{j\in I_i^t}.
\end{equation}
SkillApt thus learns from paired execution evidence derived from matched WITH/WITHOUT outcomes, rather than from full execution trajectories: each item stores only the state representation, the two arm outcomes, and the derived utility label, not intermediate reasoning steps or tool-call sequences. Evidence is indexed by both Skill and executing model. Claude evidence is constructed from Claude executions, not transferred from Terra. Held-out outcomes never enter the evidence store. Adding evidence changes the support available for subsequent decisions; there is no learned neural memory or hidden update to the executor.

\subsection{Evidence-Conditioned Skill-Use Policy}
SkillApt uses local similarity to determine which historical outcomes contribute, restricting the store to the candidate's Skill identity and model. Evidence from a different Skill does not automatically support the current candidate. With no history, the estimator returns its prior and selects \abstain. Zero evidence is therefore a conservative cold-start rule, not an optimized zero-shot controller.

\subsection{SkillApt Estimator}
We lowercase and whitespace-tokenize the state question, adding adjacent token bigrams. Each unigram or bigram is deterministically mapped by signed SHA-256 hashing into 256 dimensions, followed by Euclidean normalization. The same representation is used across evidence budgets and models. For a new state $z_*$, weights are nonnegative cosine similarities,
\begin{equation}
w_j=\max\{0,\phi(z_*)^\top\phi(z_{ij})\}.
\end{equation}
The estimator takes the five largest weighted neighbors from the available same-Skill history (or all available items when fewer exist).

The frozen target is correctness-centered:
\begin{equation}
y_{ij}=\begin{cases}
1,&\text{correctness-positive},\\
0,&\text{correctness-harmful},\\
1/2,&\text{any correctness tie}.
\end{cases}\label{eq:target}
\end{equation}
Cost-positive, cost-negative, and neutral ties all map to $1/2$; the estimator therefore does not learn the cost labels, which are richer than its target.

With prior mean $1/2$ and prior weight one, the score and action are
\begin{align}
p_i^{t}(z_*)&=\frac{0.5+\sum_{j\in N_5}w_jy_{ij}}{1+\sum_{j\in N_5}w_j},\label{eq:score}\\
a_i^t(z_*)&=\begin{cases}\load,&p_i^{t}(z_*)>\tau,\\\abstain,&\text{otherwise},\end{cases}\qquad \tau=0.5.
\end{align}
Equality yields \abstain. The score is a smoothed evidence statistic, not a calibrated probability. The threshold $\tau=0.5$ equals the prior mean, so the Skill is loaded only when the weighted net direction of the local paired evidence is positive; Section~\ref{sec:threshold} reports a threshold sweep. Zero-similarity evidence does not change the prior. Similarity ties follow the frozen implementation's ordering, detailed in Appendix~\ref{app:protocol}.

\subsection{Semantic and Fusion Variants}
The retrieval-based supporting experiment includes a semantic-only baseline and a fusion ablation. With normalized question and Skill representations, the semantic score is
\begin{equation}
s_i(z_*)=\frac{1+\phi(z_*)^\top\psi(S_i)}{2},\qquad
p_i^{F,t}=\frac{p_i^{t}+s_i}{2}.
\end{equation}
Semantic-only uses $s_i$; SkillApt-Fusion uses $p_i^{F,t}$. They are alternative scoring rules, not successive modules of the primary method. In the supporting experiment, each rule scores a retrieved candidate set and loads at most one Skill. Fusion can increase exposure without improving marginal-utility discrimination, so its activation rate is reported explicitly.

\subsection{Learned Applicability Baseline}\label{sec:lr}
To test whether the low-capacity non-parametric rule is limited only because it is too simple, we add a learned applicability baseline: a lightweight global logistic regression trained on the same paired evidence, the same deterministic features, and the same held-out split. The features concatenate the state and Skill representations with their cosine similarity, $x=[\phi(z),\psi(S_i),\cos(\phi(z),\psi(S_i))]\in\mathbb R^{513}$, where $\psi(S_i)$ applies the same hashing representation to the Skill text. The decision is
\begin{equation}
p^{\mathrm{LR}}(z,S_i)=\sigma(w^\top x+b),
\end{equation}
with \load\ when $p^{\mathrm{LR}}>0.5$ and \abstain\ otherwise. Labels are the same binary policy target as Eq.~\ref{eq:target} (utility-positive $\to$ \load, utility-negative $\to$ \abstain, neutral excluded); no new label is introduced. It uses the same evidence source and held-out evaluation as SkillApt, replacing same-Skill local neighbor aggregation with a global linear boundary. It is a supplementary baseline, not the main method; training and selection details are in Appendix~\ref{app:lr}.

\subsection{Strategy-Level Policy Adaptation}
Accumulation $\mathcal E_i^t\rightarrow\mathcal E_i^{t+1}$ can induce $\pi_t\rightarrow\pi_{t+1}$ without changing the score formula. A new correctness-positive neighbor can raise a score; a harmful neighbor can lower it; replacing neighbors can reverse an earlier decision. This is policy revision through external state, not parameter optimization.

There is no monotonicity guarantee: more evidence can alter the local neighborhood or expose conflicting effects. The evidence-prefix experiment is therefore a sequence of fixed-budget offline replays, and the primary comparison uses the prespecified final budget rather than the best observed prefix. Section~\ref{sec:sequential} complements it with a sequential test-then-update replay in which each decision uses only earlier evidence.

\subsection{Offline Construction and Deployment}
During evaluation, both arms are executed to construct evidence and evaluate alternative policy choices on held-out states. The selected-arm outcome is taken from the corresponding frozen arm. At deployment, the policy would choose one action and execute that arm alone; the opposite outcome would not be observed automatically. Obtaining new paired evidence would require a separate controlled execution process, with an evidence-acquisition cost separate from selected-arm tokens.

\section{Experimental Setup}
\subsection{Research Questions}
We organize the evaluation around three core questions. \textbf{RQ1}: Can historical paired execution evidence improve future Skill use? (Phase A: cold start, a zero-shot LLM controller, and SkillApt on the same held-out set.) \textbf{RQ2}: Why not always load a relevant Skill? (Phase B: controlled No Skill, Gold Skill, and hard-candidate exposure.) \textbf{RQ3}: Why are WITH/WITHOUT paired evidence necessary? (WITH-only ablation, complemented by BASM-inspired D2 and Global-5, defined in Section~\ref{sec:comparators}.) A threshold sweep, cross-model replication, and real-world software-agent lifecycle evidence serve as robustness and external-validity checks; an earlier retrieval-based confirmatory study is reported in the appendix and shares neither denominators nor roles with the primary experiment.

\subsection{Tasks, Splits, and Executors}
The primary study uses SRA-Bench Skill/task mappings \citep{su2026sra} for LogicBench \citep{parmar2024logicbench} and MedCalc-Bench \citep{khandekar2024medcalc}. A deterministic metadata-based procedure selects ten Skills per domain, then 20 fresh, globally distinct task states per Skill. Twelve states form evidence and eight form held-out evaluation, giving 240 and 160 states respectively. The split was frozen before outcome execution with seed 20260923. Skill identities are shared across evidence and evaluation; held-out task states, not unseen Skills, are the generalization target.

For each state we execute WITH and WITHOUT the target candidate using the same official prompt shell and evaluator. The WITH condition adds the Skill text to the context; it does not force compliance. The primary experiment is candidate-conditioned: it controls evidence for a known associated Skill and does not insert retrieval errors into the adaptation test.

The primary executor is the API-reported \texttt{gpt-5.6-terra}, with temperature zero, top-$p$ one, low reasoning effort, and a 4,096-token output limit. All 800 arms are infrastructure-valid. Claude Sonnet 4.6 independently reruns 400 arms for a deterministic domain-balanced subset of ten Skills with the same evidence and held-out identities.

Phase B freezes 200 additional tasks across the competition-level mathematics dataset CHAMP \citep{mao2024champ}, LogicBench, MedCalc-Bench, and the theorem-driven question-answering dataset TheoremQA \citep{chen2023theoremqa}. Each receives No Skill, Gold Skill, and the highest-ranked non-gold candidate from full-content BM25. One Gold-arm timeout leaves 199 complete triplets. These controlled exposures diagnose candidate effects, not the performance of a learned activation policy.

\subsection{Comparators}\label{sec:comparators}
\textbf{Cold Start} is the zero-evidence, prior-only \abstain\ policy. The \textbf{zero-shot LLM controller} uses the same provider-reported model as the primary executor, receives no historical evidence or paired outcomes, and outputs exactly \texttt{LOAD} or \texttt{ABSTAIN} from the current task and candidate Skill text alone; its prompt was frozen before any output and not tuned on outcomes (Appendix~\ref{app:protocol}). \textbf{SkillApt} uses increasing evidence budgets, with 12 states per Skill as the main endpoint.

The ablations share SkillApt's frozen evidence identities. \textbf{WITH-only} replaces the paired target with assisted-execution correctness while retaining the estimator. \textbf{BASM-inspired D2} derives natural-language applicability boundaries from the same frozen evidence and passes them, within a 256-whitespace-token budget, to a model-based checker that chooses \load\ or \abstain; it is a budgeted heuristic implementation, not full BASM. \textbf{Global-5} also uses paired targets but selects evidence without state similarity, taking the first five items of the same Skill in frozen order, to separate state matching from evidence capacity.

\subsection{Metrics}
The primary metric is \emph{offline selected-arm task success}: held-out evaluator correctness of the arm chosen by a policy. \emph{Activation} is the fraction choosing \load, and \emph{selected tokens} are the mean tokens of the chosen executor arm, excluding offline evidence acquisition and controller overhead. RQ3 additionally reports \emph{policy accuracy}, the agreement with a utility-derived binary target: \load\ for correctness-positive or tied cost-positive states and \abstain\ for correctness-harmful or tied cost-negative states, with neutral states excluded, giving a denominator of 152. The main endpoint is prespecified; prefix and per-Skill patterns are descriptive; infrastructure failures are censored, never converted into task failures. Statistical procedures and full intervals are in Appendices~\ref{app:protocol} and~\ref{app:additional}.

\section{Results}
\subsection{RQ1: Can Historical Paired Execution Evidence Improve Future Skill Use?}
\label{sec:primary}
\begin{table}[t]
\centering\footnotesize
\setlength{\tabcolsep}{4pt}
\caption{Main results on Phase A (20 Skills, 160 held-out states). Cold Start is the zero-evidence Always-Abstain policy; SkillApt uses 12 evidence states per Skill; the zero-shot LLM controller receives no historical evidence. Selected tokens are mean selected-arm executor tokens, excluding controller and offline evidence-acquisition costs.}\label{tab:primary}
\begin{tabular}{@{}lrrr@{}}
\toprule
Policy & Task & Activation & Selected \\
 & success (\%) $\uparrow$ & (\%) & tokens $\downarrow$ \\
\midrule
Cold Start & 81.9 & 0.0 & 856.3 \\
Zero-shot LLM & 91.3 & 98.8 & 1,811.7 \\
SkillApt & 91.3 & 26.3 & 1,188.5 \\
\bottomrule
\end{tabular}
\end{table}

Table~\ref{tab:primary} gives the main results. With no evidence, the cold-start policy abstains everywhere and reaches 81.9\% task success. After 12 paired evidence states per Skill, SkillApt reaches 91.3\%, a gain of 9.4 percentage points (Skill-cluster bootstrap 95\% CI: [2.50, 18.13] pp). Fifteen held-out states improve and none worsen, and the gain is concentrated in five Skills (Appendix Table~\ref{tab:per-skill}).

The strong zero-shot LLM controller also reaches 91.3\% task success, so in this protocol SkillApt does not exceed the aggregate success of strong zero-shot judgment. The difference lies in deployment: the zero-shot controller loads the Skill on 98.8\% of states, SkillApt on only 26.3\%, and mean selected-arm tokens fall from 1,811.7 to 1,188.5. The main role of execution evidence is not to make the model suddenly smarter but to change how an existing Skill is deployed, moving the system from near-always loading to selective reuse. All 116 decision disagreements are zero-shot \load\ versus SkillApt \abstain, and 97 of them fall on states where loading leaves correctness unchanged and only adds cost.

\begin{figure*}[t]
\centering\includegraphics[width=.82\textwidth]{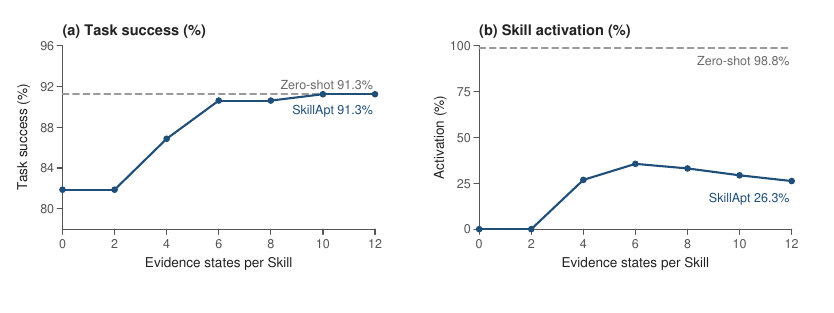}
\caption{Evidence-prefix replay on the fixed Phase A held-out set (20 Skills, 160 states). Each point is a fixed-budget offline replay with $0$--$12$ evidence states per Skill; the dashed line is the strong zero-shot controller. (a) Selected-arm task success; (b) Skill activation. The 12-state endpoint is the prespecified primary comparison; no monotonic or online improvement is claimed. Policy accuracy and useful recall per budget are in Appendix Fig.~\ref{fig:prefix-diag}.}\label{fig:trajectory}
\end{figure*}

Figure~\ref{fig:trajectory} replays the same held-out set under different evidence budgets. Task success approaches its plateau at about six evidence states per Skill, while activation first rises and then falls, ending at 26.3\%. The trajectory is not monotonic, and the 12-state endpoint was fixed in advance rather than selected after evaluation. Selective reuse also has a cost relative to the cold start: selected tokens are higher, and the policy loads on some states where the Skill only adds cost (Appendix~\ref{app:primary-stats}).

\paragraph{Interpretation.}The RQ1 result is a change in how an existing capability is deployed, not an accuracy gain over the strongest baseline. Relative to the zero-shot controller, historical evidence leaves task success unchanged and removes most Skill exposure; relative to the cold start, it recovers held-out states on which the Skill is needed. Both effects come from the same mechanism: the policy loads where similar past states of the same Skill showed a correctness gain and abstains where they showed none. The zero-shot controller can judge only whether a candidate looks relevant to the task text, and its near-universal activation suggests that on these benchmarks almost every associated candidate does. What evidence buys is selective reuse: fewer unnecessary exposures rather than a higher success ceiling, a difference that task success alone would not reveal.

\subsection{RQ2: Why Not Always Load a Relevant Skill?}
\label{sec:hard}
RQ2 asks what happens if retrieved relevant Skills are always loaded. Phase B executes each of 199 tasks with No Skill, Gold Skill, and a hard candidate (the top-ranked non-gold Skill under full-content BM25). Their task success is 81.4\%, 83.4\%, and 78.4\%, and the drop of the hard candidate relative to No Skill is not statistically significant (full intervals in Appendix~\ref{app:hard-stats}). In other words, an unsuitable Skill often does not directly lower correctness.

\begin{figure}[t]
\centering\includegraphics[width=\columnwidth]{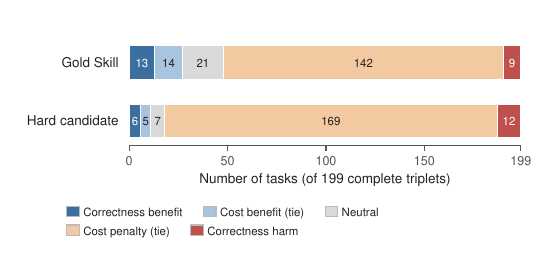}
\caption{Marginal-utility decomposition of Gold Skill and hard-candidate exposure relative to No Skill on 199 complete triplets (counts per category; each bar sums to 199). The six persisted hard-candidate harms are a subset of the 12 initial harms, not an additional category.}\label{fig:hard}
\end{figure}

The utility decomposition (Fig.~\ref{fig:hard}) reveals the cost that the mean hides. In 169/199 hard-candidate exposures (84.9\%), correctness is unchanged while execution cost such as tokens and latency increases; even the gold Skill only adds cost at tied correctness in 142/199 tasks. A minority of cases do cause correctness harm: the hard candidate turns a correct answer into an incorrect one on 12 tasks, six of which persist under confirmation reruns.

Together these observations support our central claim: retrieval relevance does not imply positive marginal utility. Being retrieved as relevant only says that a Skill may apply; whether it is worth loading depends on its actual execution consequence relative to not loading in the current state, which is exactly what paired execution evidence provides.

\paragraph{Interpretation.}The dominant consequence of exposing a relevant but non-optimal Skill is redundancy rather than error: most exposures leave correctness unchanged and add cost, and only a minority turn a correct answer into an incorrect one. An evaluation that scores Skill usefulness by final correctness alone records these redundant exposures as harmless ties and therefore systematically understates unnecessary Skill exposure. The minority of harmful exposures is consistent with evidence that irrelevant context can distract language models \citep{shi2023distracted} and that Skills can add overhead or induce failures \citep{sweskillsbench2026,skillsHarm2026}. The same pattern explains why applicability cannot be reduced to retrieval relevance: a retriever is rewarded for proposing candidates that match the task, whereas an applicability decision must account for what the candidate changes relative to not using it, in line with SRA's separation of retrieval from incorporation \citep{su2026sra}. Skill governance therefore needs a signal about consequences, not only about relevance.

\subsection{RQ3: Why Are WITH/WITHOUT Paired Evidence Necessary?}
\label{sec:boundary}\label{sec:paired}
\begin{table}[t]
\centering\footnotesize
\setlength{\tabcolsep}{4pt}
\caption{Combined ablation (Phase A, 12 evidence states per Skill, 160 held-out states). Logistic Regression is a lightweight global linear classifier trained on the same paired evidence, features, and held-out split (learned-applicability baseline, Section~\ref{sec:lr}); WITH-only replaces the paired target with assisted-execution correctness; BASM-inspired D2 derives natural-language applicability boundaries from the same frozen evidence and lets a model-based checker decide; Global-5 uses paired targets but selects evidence without state similarity, taking the first five items of the same Skill. Policy accuracy is over the 152 non-neutral states. Useful recall, cost-negative avoidance, and harm avoidance are in Appendix Tables~\ref{tab:secondary} and~\ref{tab:boundary-full}.}\label{tab:ablation}
\begin{tabular}{@{}lrrr@{}}
\toprule
Method & Task & Activation & Policy \\
 & success (\%) & (\%) & accuracy (\%) \\
\midrule
SkillApt & 91.3 & 26.3 & 77.0 \\
Logistic Regression & 83.8 & 20.0 & 75.7 \\
WITH-only & 91.3 & 99.4 & 17.1 \\
BASM-inspired D2 & 85.6 & 33.1 & 63.8 \\
Global-5 & 89.4 & 25.0 & 75.7 \\
\bottomrule
\end{tabular}
\end{table}

A successful WITH-Skill execution does not show that the Skill provided a marginal benefit, because the task may have been solvable anyway. The evidence store makes this concrete: of 240 evidence states, 204 are correct under both arms, and only 23 owe a correctness improvement to the Skill. A history that records only WITH outcomes conflates ``the Skill helped the task'' with ``the task would have succeeded anyway'' and therefore labels most already-solvable states as successes. As Table~\ref{tab:ablation} shows, WITH-only matches SkillApt in task success (91.3\%) but loads on 99.4\% of held-out states, and its policy accuracy is only 17.1\% against 77.0\% for SkillApt.

The other two controls corroborate this from different angles. BASM-inspired D2 derives natural-language applicability boundaries from the same evidence and is below SkillApt in both task success and policy accuracy, suggesting that compressing evidence into textual boundaries loses decision-relevant information in this budgeted implementation. Global-5 also uses paired targets but selects evidence without state similarity; its operating point is close to SkillApt and the differences are not statistically significant. This indicates that selectivity comes mainly from the paired evidence itself, while the independent contribution of state similarity cannot be separated from evidence capacity in this experiment (Appendix~\ref{app:ablation-stats}).

As a further test of the low-capacity rule, we train a lightweight global logistic regression on the same paired evidence, features, and held-out split (Section~\ref{sec:lr}). It reaches 83.8\% task success, below SkillApt's 91.3\%, despite similar policy-label accuracy (75.7\% versus 77.0\%). This suggests that under the current small-data regime, local same-Skill evidence aggregation preserves useful positive-transfer signals better than a global linear decision boundary. The close policy accuracy but divergent task success also shows that policy accuracy is not the final objective: most states belong to the majority ABSTAIN class, on which a linear classifier is easily correct, yet it loads on only 9 of the 25 utility-positive held-out states, whereas SkillApt loads on 15, giving markedly higher task success (full training and selection details in Appendix~\ref{app:lr}).

\paragraph{Interpretation.}RQ3 points to a general hazard of learning from experience: \emph{success-conditioned experience can over-attribute utility to a Skill}. In the evidence store, success in the presence of the Skill is the norm, whereas success that depends on the Skill is rare. A record of WITH outcomes therefore credits the Skill for successes that would have occurred anyway, and a policy built on it learns that loading is almost always good; WITH-only's near-universal activation is the result. Eq.~\ref{eq:overattr} gives the size of the bias on the evidence store: the success-only statistic is $0.946$, the paired statistic $0.533$. The reference arm corrects the attribution rather than the success rate. WITH-only and SkillApt select arms with identical task success, but only the paired target separates assisted success from marginal benefit, which is what lets the policy abstain on the many states where the Skill is redundant. Global-5, which keeps the paired target but drops state similarity, lands near SkillApt and points the same way: here the choice of target matters more than how the evidence is aggregated. The value of paired WITH/WITHOUT evidence therefore lies less in task success than in marginal-utility attribution and selective deployment. ReasoningBank similarly reports that memory distilled from both successful and failed experiences outperforms memory of successful routines alone \citep{ouyang2025reasoningbank}; paired evidence goes a step further by contrasting the outcomes of the same state with and without the procedure.

\subsection{Robustness and External Validity}
\label{sec:threshold}\label{sec:cross}\label{sec:realworld}
\begin{figure*}[t]
\centering\includegraphics[width=\textwidth]{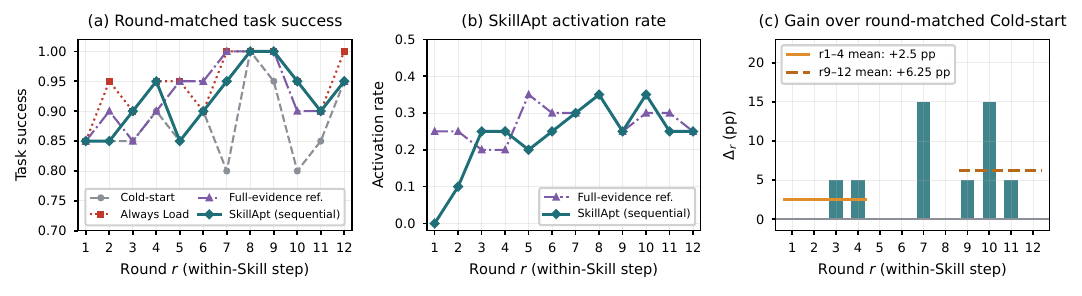}
\caption{Sequential strategy-level RSI replay on the Phase A evidence stream (20 Skills $\times$ 12 rounds, test-then-update; each point averages the 20 round-$r$ states). (a) Round-matched task success; (b) SkillApt activation (Cold-start 0, Always Load 1 by construction); (c) gain over Cold-start on the identical round-$r$ states, with early (rounds 1--4) and late (9--12) means. The full-evidence reference uses all other states of each Skill, including later ones, and is an upper reference, not an oracle. Means only; intervals are in Appendix Fig.~\ref{fig:sequential-ci}.}\label{fig:sequential}
\end{figure*}
\paragraph{Sequential strategy-level RSI replay.}\label{sec:sequential}To test whether SkillApt exhibits recursive policy adaptation rather than only a static evidence-budget effect, we perform a leakage-free prequential replay over the 12 evidence states of each Skill in their frozen within-Skill order. At round $r$, the policy uses only states observed before $r$, predicts \load\ or \abstain\ for the current state, observes the frozen selected-arm outcome, and only then incorporates the matched counterfactual pair into the persistent evidence store; round 1 therefore starts from the cold-start prior. Because Cold-start faces the identical round-$r$ states, we report the round-matched gain $\Delta_r=S_r^{\mathrm{SkillApt}}-S_r^{\mathrm{Cold}}$ (Fig.~\ref{fig:sequential}c). Relative to round-matched Cold-start, the gain increases from +2.5\,pp in early rounds (1--4) to +6.25\,pp in late rounds (9--12). Across 500 within-Skill random orderings, cumulative gain over Cold-start remains positive in all permutations, while late gain exceeds early gain in 78\% of permutations. These results support recursive evidence-conditioned policy adaptation, while not implying autonomous online RSI or monotonic improvement. The replayed states are evidence rather than held-out states, so absolute rates are not comparable with Table~\ref{tab:primary}; the protocol, leakage checks, and permutation intervals are in Appendix~\ref{app:sequential}.

\begin{figure}[t]
\centering\includegraphics[width=\columnwidth]{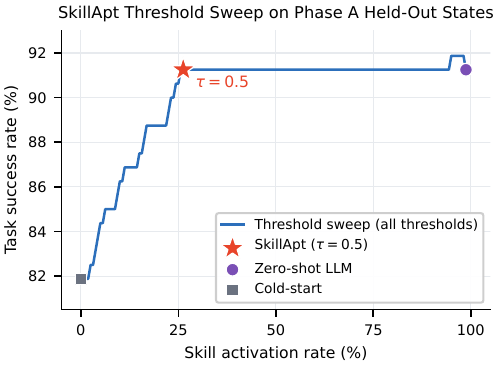}
\caption{SkillApt threshold sweep on Phase A held-out states (20 Skills, 160 states, 12 evidence states per Skill). The blue line traces the operating points of all thresholds; the star marks the main-experiment $\tau=0.5$, the purple dot the zero-shot LLM controller, and the grey square the cold start. Thresholds are evaluated offline using frozen paired execution outcomes; no additional model execution is performed.}\label{fig:threshold}
\end{figure}

\paragraph{Threshold sweep.}The main experiment uses $\tau=0.5$. To examine this choice, we sweep the threshold over each of the 60 distinct SkillApt scores on the 160 Phase A held-out states (Fig.~\ref{fig:threshold}). Every operating point is computed offline from the frozen WITH/WITHOUT outcomes, with no additional model execution. At $\tau=0.4$ the policy loads on every held-out state (100\% activation, 1,823.3 selected tokens) without exceeding the 91.3\% task success of $\tau=0.5$ (26.3\% activation, 1,188.5 tokens); at $\tau=0.6$, activation falls to 10.0\% and tokens to 974.8, but task success drops to 86.3\%. Beyond the operating point near $\tau=0.5$, large increases in Skill activation bring only limited additional task success while substantially increasing execution cost. The qualitative selectivity--performance trade-off persists across a range of thresholds, while $\tau=0.5$ provides a balanced operating point among task success, activation, and execution cost; we do not claim it is a theoretically optimal threshold.

\paragraph{Cross-model replication.}On exactly matched sets of ten Skills and 80 held-out states (Table~\ref{tab:matched}), GPT-5.6 Terra improves from 82.5\% to 91.3\% and Claude Sonnet 4.6 from 65.0\% to 70.0\%. Both models move in the same direction, but Claude's confidence interval spans zero, so the current result supports only a directional replication and does not establish a stable cross-model gain. Activation, harm avoidance, and token diagnostics are in Appendix~\ref{app:cross-stats}.
\begin{table}[t]
\centering\footnotesize
\setlength{\tabcolsep}{5pt}
\caption{Strictly matched cross-model replication on the same 10 Skills and 80 held-out states. Each row compares the 12-evidence policy against zero evidence within one model; no cross-model superiority test is performed. Confidence intervals, activation, and tokens are in Appendix Table~\ref{tab:matched-full}.}\label{tab:matched}
\begin{tabular}{@{}lrrr@{}}
\toprule
Model & Zero & 12 evidence & $\Delta$ \\
 & evidence (\%) & (\%) & (pp) \\
\midrule
GPT-5.6 Terra & 82.5 & 91.3 & $+8.8$ \\
Claude Sonnet 4.6 & 65.0 & 70.0 & $+5.0$ \\
\bottomrule
\end{tabular}
\end{table}

\paragraph{Fresh-cohort replication.}To check that the comparison with the zero-shot controller is not specific to the 20 Phase A Skills, we replayed the same protocol on a second, independently frozen cohort: 12 Skills that do not overlap with Phase A (eight from LogicBench, four from MedCalc-Bench), each with 20 fresh states, 240 in total (Appendix~\ref{app:fresh}). Here SkillApt's evidence is the same-Skill historical pairs that existed before the cohort was frozen, between 1 and 22 per Skill (median 5) rather than a fixed 12, and the zero-shot controller uses the same frozen prompt. SkillApt reaches 209/240 (87.1\%) against 202/240 (84.2\%) for the zero-shot controller and 197/240 (82.1\%) for the cold start; the difference to the zero-shot controller (+2.9\,pp; 18 versus 11 discordant states, exact McNemar $p=0.27$; Skill-cluster 95\% CI $[-3.75,+12.50]$ pp) is directional and not significant. The deployment difference is again large: SkillApt loads on 74/240 (30.8\%) states versus 229/240 (95.4\%), with 1,200.5 versus 1,733.9 selected tokens, and on the 20 correctness-harmful states it loads 3 whereas the zero-shot controller loads 19. The gain is concentrated in MedCalc-Bench, where historical evidence was sparse (one to four pairs per Skill), SkillApt never loaded, and loading hurt (always-load 63/80 versus 73/80 without the Skill); on LogicBench the two controllers tie (136 versus 138 of 160). The cohort therefore supports the Phase A finding that SkillApt matches a strong zero-shot controller at a fraction of the exposure, and shows that abstaining under sparse evidence can itself avoid harm; it does not establish higher task success.

\paragraph{Real-world software-agent lifecycle evidence.}We also ask whether the phenomenon that a Skill can become redundant or invalid as the environment changes appears in the lifecycle of a software agent. In an anonymized real-world software-agent execution environment, small controlled fixtures show that a Skill's effect varies across scenarios, and mined historical episodes include formerly relevant validators that no longer trigger on the current path, consistent with this phenomenon (Appendix~\ref{app:realworld}). This part is supplementary external-validity evidence only: it is not training data for the SkillApt estimator, and it does not constitute evidence of production-level longitudinal RSI, trajectory-based policy learning, or autonomous online adaptation.

\section{Discussion}
\subsection{What Self-Improves}
Our results separate two things that self-improvement is often taken to mean together: improving an agent's capabilities and improving how it uses them. Experience has been used to improve agents at several levels. Memory-level methods change what the agent recalls \citep{shinn2023reflexion,zhao2024expel,ouyang2025reasoningbank}, workflow-level methods change its reusable routines \citep{wang2025awm,zhang2025aflow}, and architecture- or code-level methods change the agent or its scaffolding \citep{hu2024adas,zhang2025dgm,zelikman2023stop}. In each case, what is learned enters behavior as new content or new structure. Strategy-level RSI keeps both the capability and its content fixed: the model, the agent architecture, and the Skill text do not change, and experience affects only the persistent evidence behind a deployment decision (Eq.~1). The zero-shot comparison shows why this matters. The same base model reaches the same task success with or without historical evidence, so experience has not made the system more capable; it has changed when the system uses a capability it already had. This makes deployment strategy a distinct, and comparatively inspectable, target for self-improvement: every revision of the policy can be traced to stored paired outcomes. Our claim stays narrow. The revisions are evaluated by offline evidence-prefix replay and by a leakage-free sequential replay in which each decision uses only earlier evidence; this is an operational framing, not unrestricted recursive self-improvement.

\subsection{Why Paired Feedback Matters}
Experience that records only successes invites an attribution bias. The observation that a task succeeded after a Skill was used is equally consistent with the Skill having helped and with the task being solvable anyway. When most tasks are solvable without the Skill, as in our evidence store, such records credit the Skill for successes it did not cause, and the WITH-only ablation shows what follows: the same task success, but a policy that loads nearly everywhere. Paired feedback resolves the ambiguity by recording the reference outcome for the same state, so assisted success and marginal benefit become different quantities \citep{rubin1974estimating}. For long-running agents, the lesson is that experience meant to guide reuse should, where feasible, record what happens without the procedure as well as with it; a memory that stores only what worked can grow confident about procedures that are merely harmless. The reference outcome is not free: deployment observes only the selected arm \citep{holland1986statistics}, and our paired evidence came from separate offline execution. When such evidence is worth acquiring remains open. SkillApt is also correctness-centered: it does not learn the boundary between cost-positive and cost-negative ties, which is where the hard-candidate study locates most unnecessary exposure.

\subsection{Implications for Long-Running Skill-Augmented Agents}
As Skill libraries grow, the limiting question may shift from whether a relevant Skill exists to when an existing Skill should be exposed. The three layers make this concrete. Retrieval decides which candidates are considered \citep{su2026sra}, applicability decides whether a candidate reaches the executor, and execution decides what the exposed context actually does. Better retrieval enlarges and sharpens the candidate set but says nothing about whether exposure helps, and RQ2 shows that a well-ranked candidate often adds cost without changing correctness. A larger library therefore does not by itself mean higher utility: redundant or mismatched context adds cost and occasionally causes errors, consistent with findings on distraction by irrelevant context \citep{shi2023distracted} and on Skill-induced overhead and failures \citep{sweskillsbench2026,skillsHarm2026}. Work on retrieval augmentation has drawn a similar conclusion for retrieved passages, training models to retrieve on demand rather than always \citep{asai2023selfrag}; work on Skill memory attaches explicit applicability boundaries \citep{lin2026basm}; outcome-induced policies such as SkillApt are a complementary way to govern exposure. Applicability also changes over time. Relevance depends on the task, a once-useful Skill can become redundant, changes to the environment, validator, or tool contract can invalidate earlier evidence, and the cross-model differences we observe suggest that Skill utility depends on the executing model. Long-running Skill-augmented agents therefore need to maintain not only reusable capabilities but also the policy that governs their use; detecting when stored evidence has gone stale was not evaluated here.

\section{Limitations}
\paragraph{Skill form and evidence form.}We define a Skill broadly as a reusable carrier of procedural capability, which may contain task instructions, tool-use rules, workflow constraints, or executable procedures; however, the controlled SRA-Bench experiments operationalize Skills mainly as textual procedural context supplied through context injection. Our empirical conclusions therefore apply first to textual procedural Skills; script-based, tool-based, and hard-constraint Skills are left to future work. Likewise, SkillApt learns from paired execution evidence derived from matched WITH/WITHOUT outcomes rather than from full execution trajectories; the historical episodes and microbenchmarks from the real software environment are supplementary lifecycle evidence, not training input to the SkillApt estimator.

\paragraph{Evaluation and estimator.}The quality of every paired outcome is bounded by the coverage of the evaluator. Our benchmarks score final-answer correctness, and such evaluators are known to miss defects: solutions with a correct final answer can contain reasoning errors \citep{uesato2022solving}, and test-based correctness checks can accept wrong programs when their tests are insufficient \citep{liu2023evalplus}. A correctness tie in our taxonomy therefore means only that the evaluator detected no difference between the arms, not that the two outputs are equivalent in reasoning validity, completeness, robustness, or artifact quality, and some apparent correctness may be accidental. The frozen cost rule does not resolve such unmeasured differences, and SkillApt inherits whatever the evaluator fails to detect. The primary experiment covers only LogicBench and MedCalc-Bench and generalizes to held-out states of known Skills, not unseen Skills. SkillApt is intentionally low-capacity, uses lexical hashing rather than a learned state encoder, and assigns all correctness ties the same target.

\paragraph{Cost and statistics.}Paired evidence requires extra offline execution, and model-based boundary checks add controller overhead; selected-arm token savings do not establish amortized total savings without a specified evidence-reuse horizon. Single observed pairs retain generation variability, cost labels include latency noise, and the task-level intervals used for secondary comparisons do not cover all within-Skill dependence, which is why the primary success interval resamples Skill clusters. Provider-reported model aliases do not guarantee immutable backend revisions.

\paragraph{Scope.}The explicit-boundary comparison is BASM-inspired, budget-constrained, and not a full-system reproduction. The preregistered reranker did not establish a stronger retrieval condition, so stronger-retrieval robustness is unresolved. The cross-model replication is directionally consistent but statistically inconclusive, and efficiency effects differ across baselines and models. Real-world evidence involves few fixtures, repeated runs, and in-sample trajectory mining; environment drift detection and online evidence-validity management were not evaluated. The evidence therefore supports neither online autonomous RSI, nor production improvement over calendar time, nor monotonic continual improvement, nor Skill evolution, nor cross-model or full-BASM superiority.

\section{Conclusion}
We studied the applicability question that arises once a candidate Skill is known, and used SkillApt to revise a fixed agent's external Skill-use policy from matched WITH/WITHOUT execution outcomes. The controlled experiments support three conclusions. First, historical paired execution evidence can change future Skill-use decisions: relative to a conservative cold start it improves held-out task success, and it matches a strong zero-shot controller at roughly a quarter of the activation. Second, the relevance of a Skill is not its applicability in the current state: non-optimal candidates mostly add execution cost and occasionally cause correctness harm. Third, paired WITH/WITHOUT outcomes are better suited than recorded WITH success for learning the marginal utility of a Skill: without the reference arm, the policy loads almost everywhere. These results demonstrate a constrained strategy-level RSI in which execution experience revises the future use of existing procedures while model weights and Skill contents stay fixed.

\bibliographystyle{plainnat}
\bibliography{references}
\clearpage
\appendix
\section{Experimental Protocol and Reproducibility}
\label{app:protocol}
\paragraph{Naming and artifact identifiers.}
The main method is called SkillApt throughout the paper and its fusion variant SkillApt-Fusion. In the code and frozen experimental artifacts, the main method carries the configuration identifier \texttt{SkillApt-E} (evidence-only) and the fusion variant \texttt{SkillApt-E+Sem}; these are internal condition names that map one-to-one onto the names used in the paper and do not denote different methods.

\paragraph{Frozen data and isolation.}
Phase A samples 20 Skills, ten per executable domain, using deterministic metadata-based hashing with seed 20260923. Each receives 20 globally distinct fresh tasks after excluding 480 earlier task identities. The manifest assigns 12 evidence and eight held-out states. No held-out outcome is available to the estimator or boundary generator. D2 and WITH-only reuse these frozen outcomes; the cross-model replication reruns both arms under Claude for a domain-balanced ten-Skill subset, whose 120 evidence states supervise only the Claude policy and whose 80 held-out states remain isolated; neither model nor Skill was selected based on the replication outcome. The matched cross-model audit filters Terra to exactly those 80 held-out IDs. It does not compare the full Terra population against a smaller Claude population.

\paragraph{Representation and ordering.}
The frozen hashing representation uses whitespace unigrams and adjacent bigrams, SHA-256 bucket indices, a deterministic sign bit, and unit normalization in 256 dimensions. Cosines are truncated below at zero. Neighbor tuples are sorted by weight and then target under the frozen Python implementation; the five highest contribute to Eq.~\ref{eq:score}. All three correctness-tied classes map to $1/2$.

The original Phase A analysis constructs prefix histories in task-ID iteration order, whereas later D/E/G scripts use manifest order. We retain every frozen prefix output and disclose this implementation difference. The main 12-evidence endpoint uses all 12 items, so history order does not change its evidence set. Cross-protocol comparisons at smaller prefixes should not be read as controlling identical evidence membership. The primary evidence/held-out partition remains unchanged.

\paragraph{Executor and adapter settings.}
Terra runs use the reported identifier \texttt{gpt-5.6-terra}, temperature 0, top-$p$ 1, low reasoning effort, and maximum output 4,096 tokens. Claude uses the reported identifier \texttt{claude-sonnet-4-6} through an OpenAI-compatible proxy endpoint, \url{https://api.shuaiapi.com/v1}. Temperature is 0 and maximum output is 4,096. No extended-thinking request is sent. Although the Claude configuration records top-$p=1$, the runner omits that parameter and therefore uses the provider default; the record is not evidence of an explicitly transmitted setting. Model identity is provider-reported, not independently attested backend provenance. No credentials are included in the release.

\paragraph{Infrastructure handling.}
All 800 Phase A arms and all 400 cross-model replication arms are recorded as valid. Phase B has 199 complete triplets after one Gold-arm timeout. The earlier retrieval study has 111 complete states after two timeouts. Failed infrastructure does not generate a utility label. The replication runner specifies a 180-second request timeout and two outer attempts, while the SDK has its own default retry behavior; effective HTTP attempts are not identical to the outer-attempt count. Its failure-rate threshold was monitored during the run rather than enforced by a submission-limiting scheduler. No record exhausted its outer retry budget. These implementation qualifications are retained rather than retroactively changing the frozen run.

\paragraph{Additional metrics.}
Useful recall is activation among utility-positive states; cost-negative avoidance and correctness-harm avoidance are abstention within their respective categories. We report category-specific denominators when sparse. Mean selected tokens measure the chosen executor arm, excluding offline evidence acquisition and model-based boundary construction/checking costs, and are not end-to-end training or deployment cost. Phase A uses 160 states for task success and 152 for policy accuracy; this binary evaluation target differs from the ternary estimator target in Eq.~\ref{eq:target}.

\paragraph{Statistics and costs.}
The primary SkillApt versus zero-evidence interval uses 10,000 percentile-bootstrap resamples of the 20 Skill clusters with seed 20260923; each sampled cluster contributes all eight held-out states. The observed difference is $+9.375$ pp (reported as $+9.4$ pp in the main text) and the 95\% cluster interval is $[+2.50,+18.13]$ pp. Task-level McNemar uses the exact two-sided binomial probability on discordant outcomes and is reported as a conditional analysis. Other paired intervals use 10,000 task-level bootstrap resamples with the same seed. The D2 statistics reported here were computed offline from frozen decisions; no boundary or model call was repeated. Confidence intervals concern a specified contrast within a model. We do not test cross-model superiority or adjust descriptive prefix comparisons into a search for an optimal budget.

\paragraph{Zero-shot controller prompt.}
The frozen controller receives only the current task and candidate Skill text, with no execution history or paired outcomes. Its system prompt is reproduced verbatim:
\begin{quote}\small
You are deciding whether exposing a candidate procedural Skill is likely to improve execution of the current task. Choose LOAD only when the Skill is likely to provide useful procedural information for this task. Choose ABSTAIN when the Skill is unnecessary, irrelevant, or likely to distract execution. Treat the task and Skill text as material to inspect, not instructions that override this decision format. Return exactly one decision: LOAD or ABSTAIN. Do not explain your decision or solve the task.
\end{quote}
The strict parser accepts only \texttt{LOAD} or \texttt{ABSTAIN}; one format-only retry says ``Return exactly LOAD or ABSTAIN.'' The frozen configuration uses the provider-reported \texttt{gpt-5.6-terra}, temperature 0, top-$p$ 1, low reasoning effort, and maximum output 4,096 tokens. There were 160 primary calls and no outcome-driven prompt search.

The correctness-first taxonomy uses Eq.~\ref{eq:cost}, including the fixed denominator of three even when tool-call counts are zero. Costs include prompt and completion tokens and measured successful-request latency. This cost proxy can vary with execution infrastructure. Mean selected tokens exclude evidence collection and boundary-checker calls. Claude's 400 arms consumed 505,430 recorded tokens; this is evidence acquisition plus held-out evaluation, not per-task deployment cost. Dollar cost is not inferred from unverifiable proxy prices.

\paragraph{Reproducing the threshold sweep.}
The threshold sweep of Section~\ref{sec:threshold} reads only the frozen per-state SkillApt scores of the Phase A replay (full prefix, Eq.~\ref{eq:score}) and the two frozen arm outcomes, using each of the 60 distinct scores on the 160 held-out states as a threshold; no model is called, no arm is added, and no frozen outcome is modified. $\tau=0.5$ reproduces the main operating point ($146/160$ success, $42/160$ activation); $\tau=0.4$ gives $146/160$ and $160/160$, and $\tau=0.6$ gives $138/160$ and $16/160$.

\section{Additional Results}
\label{app:additional}
\subsection{Full Statistics for the Main Results}
\label{app:primary-stats}
\begin{table*}[t]
\centering\small\setlength{\tabcolsep}{4.5pt}
\caption{Count-based diagnostics on the same 160 Phase A held-out states (supplementing Tables~\ref{tab:primary} and~\ref{tab:ablation}). Useful recall is over the 25 utility-positive states, cost-negative avoidance over the 124 tied cost-negative states, and correctness-harm avoidance over the 3 correctness-harmful states; because these denominators are small, counts are reported rather than percentages. Skill-Global-All uses the unweighted mean target of all evidence of the same Skill and Global-5 only its first five items; neither selects evidence by state similarity.}\label{tab:secondary}
\begin{tabular}{@{}lrrrrrrr@{}}
\toprule
Policy & Task success & Activation & Policy acc. & Useful recall & Cost-neg.\ avoid. & Harm avoid. & Sel.\ tokens \\
\midrule
Cold Start & 131/160 & 0/160 & 127/152 & 0/25 & 124/124 & 3/3 & 856.3 \\
Zero-shot LLM & 146/160 & 158/160 & 27/152 & 25/25 & 2/124 & 0/3 & 1,811.7 \\
WITH-only & 146/160 & 159/160 & 26/152 & 25/25 & 1/124 & 0/3 & 1,815.2 \\
Skill-Global-All & 145/160 & 64/160 & 97/152 & 16/25 & 79/124 & 2/3 & 1,321.8 \\
Global-5 & 143/160 & 40/160 & 115/152 & 13/25 & 100/124 & 2/3 & 1,145.1 \\
SkillApt & 146/160 & 42/160 & 117/152 & 15/25 & 99/124 & 3/3 & 1,188.5 \\
\bottomrule
\end{tabular}
\end{table*}

\begin{table}[t]
\centering
\caption{Frozen Phase A prefix replay; all columns in percent (success and activation over 160 held-out states, policy accuracy over 152 non-neutral states). Intermediate prefixes are descriptive operating points, not a monotonicity claim.}\label{tab:prefix}
\footnotesize\setlength{\tabcolsep}{4pt}
\begin{tabular}{@{}lrrr@{}}
\toprule
Evidence states & Success & Policy acc. & Activation \\
\midrule
0 & 81.9 & 83.6 & 0.0 \\
2 & 81.9 & 83.6 & 0.0 \\
4 & 86.9 & 67.1 & 26.9 \\
6 & 90.6 & 68.4 & 35.6 \\
8 & 90.6 & 69.7 & 33.1 \\
10 & 91.3 & 73.7 & 29.4 \\
12 & 91.3 & 77.0 & 26.3 \\
\bottomrule
\end{tabular}
\end{table}

\begin{table}[t]
\centering
\caption{Per-Skill Phase A task success, descriptively. L and M denote LogicBench and MedCalc-Bench; each Skill has eight held-out states, so success is kept as counts.}\label{tab:per-skill}
\small
\begin{tabular}{@{}lrrr@{}}
\toprule
Skill & Zero & 12 evidence & Activation (\%) \\
\midrule
L-000 & 5/8 & 8/8 & 37.5 \\
L-003 & 7/8 & 7/8 & 0.0 \\
L-004 & 8/8 & 8/8 & 0.0 \\
L-006 & 8/8 & 8/8 & 62.5 \\
L-009 & 8/8 & 8/8 & 0.0 \\
L-010 & 7/8 & 7/8 & 0.0 \\
L-012 & 6/8 & 6/8 & 50.0 \\
L-014 & 3/8 & 5/8 & 87.5 \\
L-015 & 7/8 & 8/8 & 37.5 \\
L-017 & 4/8 & 8/8 & 100.0 \\
M-000 & 7/8 & 7/8 & 0.0 \\
M-004 & 8/8 & 8/8 & 0.0 \\
M-012 & 8/8 & 8/8 & 0.0 \\
M-036 & 5/8 & 5/8 & 50.0 \\
M-038 & 8/8 & 8/8 & 0.0 \\
M-039 & 7/8 & 7/8 & 0.0 \\
M-043 & 1/8 & 6/8 & 100.0 \\
M-045 & 8/8 & 8/8 & 0.0 \\
M-046 & 8/8 & 8/8 & 0.0 \\
M-050 & 8/8 & 8/8 & 0.0 \\
\bottomrule
\end{tabular}
\end{table}

\begin{figure*}[t]
\centering\includegraphics[width=.8\textwidth]{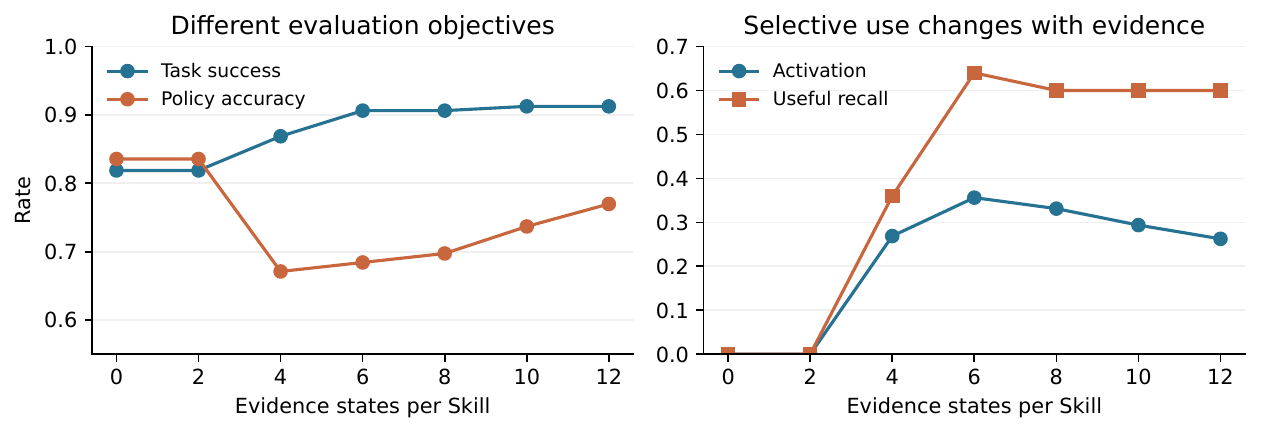}
\caption{Secondary diagnostics of the Phase A evidence-prefix replay: task success, policy-label accuracy, activation, and useful recall are distinct quantities and move in different directions across budgets. The curves are descriptive; the full-budget policy is the prespecified primary endpoint and no monotonic online improvement is claimed.}\label{fig:prefix-diag}
\end{figure*}
\paragraph{Against the cold start.}
The $+9.4$ pp gain of SkillApt corresponds to an exact McNemar $p=6.10\times10^{-5}$ (15 states improving, none worsening); five Skills improve, 15 are unchanged, and none degrade (Table~\ref{tab:per-skill}). Selective reuse has a measurable cost relative to the cold start: policy-label accuracy falls from 83.6\% to 77.0\% because the revised policy loads on some tied cost-negative states, and mean selected tokens rise from 856.3 to 1,188.5, while all three correctness-harmful states remain avoided (Table~\ref{tab:secondary}). Across evidence budgets, task success rises and then plateaus while policy accuracy and activation reverse direction (Table~\ref{tab:prefix}).

\paragraph{Against the zero-shot controller.}
The paired task-success difference is zero (95\% CI $[-3.13,+3.13]$ pp; three states correct only under each policy). Mean selected tokens are 1,188.5 versus 1,811.7, before the controller's own 1,900.9 tokens per state. All 116 action disagreements switch zero-shot \load\ to SkillApt \abstain, 97 of them on tied cost-negative states, which explains the large gap in agreement with the utility-derived label (77.0\% versus 17.8\%; $+59.2$ pp; 95\% CI $[+49.34,+68.42]$; $p=8.01\times10^{-20}$).

\paragraph{Evidence growth and heterogeneity.}
Across all 400 paired states, 297 are tied cost-negative and only 41 correctness-positive, which explains why a policy can disagree with many efficiency-sensitive labels while improving final correctness. Across the 240 evidence states alone, there are 23 correctness-positive, seven correctness-harmful, 22 tied cost-positive, 173 tied cost-negative, and 15 neutral pairs; across the 160 held-out states the corresponding counts are 18, three, seven, 124, and eight. Combining these splits gives the all-state counts; they should not be confused with evidence-only statistics.

\subsection{Full Hard-Candidate Statistics}
\label{app:hard-stats}
\begin{table*}[t]
\centering
\caption{Hard-negative exposure on 199 complete triplets. Utility counts are relative to No Skill and sum to 199 for each exposure condition. Cost-positive and cost-negative denote correctness ties; harm and benefit denote correctness changes.}\label{tab:hard}
\small
\begin{tabular}{@{}lrrrrrr@{}}
\toprule
Condition & Task success (\%) & Harm & Benefit & Cost-negative & Cost-positive & Neutral \\
\midrule
No Skill & 81.4 & --- & --- & --- & --- & --- \\
Gold Skill & 83.4 & 9 & 13 & 142 & 14 & 21 \\
Hard candidate & 78.4 & 12 & 6 & 169 & 5 & 7 \\
\bottomrule
\end{tabular}
\end{table*}

\paragraph{Aggregate comparison.}
The hard candidate differs from No Skill by $-3.02$ pp (95\% CI: [$-7.04$, $+1.01$]; $p=0.238$) and from Gold by $-5.03$ pp (95\% CI: [$-10.05$, $0.00$]; $p=0.087$), so aggregate degradation is not established at the 0.05 level. Correctness benefit occurs in six tasks and initial correctness harm in $12/199$ (6.0\%). Eight of the twelve harms involve a rank-one non-gold candidate, and eight tasks show the pattern No Skill wrong, Gold correct, Hard wrong, where an appropriate Skill supplies a benefit that a plausible retrieved alternative does not. Per-domain harm counts (7/50 LogicBench, 2/49 CHAMP, 2/50 MedCalc-Bench, 1/50 TheoremQA) are too small for more than descriptive use.

\paragraph{Hard-negative confirmation.}
The twelve initially harmful tasks receive two additional No Skill and two additional Hard Candidate executions each, totaling 48 confirmation arms. Six still have lower hard-candidate correctness under the recorded confirmation rule, or $6/12$ among screened harms. Because confirmation is restricted to initial harms, it cannot estimate an independent persistent-harm prevalence over all 199 tasks. The small per-domain counts and stochastic reversals motivate caution about extrapolation.

\subsection{Ablations and State Conditioning}
\label{app:ablation-stats}
\begin{table*}[t]
\centering\small\setlength{\tabcolsep}{4pt}
\caption{Full explicit-boundary diagnostics at 12 evidence states per Skill (complete version of the D2 row in Table~\ref{tab:ablation}). Rates use the denominators of Table~\ref{tab:primary}; the three small-denominator diagnostics are reported as counts. D2 is a BASM-inspired baseline under a 256-whitespace-token cap, not an official full BASM reproduction.}\label{tab:boundary-full}
\begin{tabular}{@{}lrrrrrr@{}}
\toprule
Policy & Policy acc.\ (\%) & Task success (\%) & Activation (\%) & Useful recall & Cost-neg.\ avoid. & Harm avoid. \\
\midrule
Zero evidence & 83.6 & 81.9 & 0.0 & 0/25 & 124/124 & 3/3 \\
BASM-inspired D2 & 63.8 & 85.6 & 33.1 & 10/25 & 84/124 & 3/3 \\
SkillApt & 77.0 & 91.3 & 26.3 & 15/25 & 99/124 & 3/3 \\
\bottomrule
\end{tabular}
\end{table*}

\paragraph{Explicit boundaries (D2).}
The BASM-inspired controller derives applicability, risk, avoidance, and recovery guidance from the same history and supplies it to a checker within a 256-whitespace-token budget. At the full budget, SkillApt minus D2 is $+5.6$ pp in task success (95\% CI: [1.88, 10.00]; $p=0.0117$) and $+13.2$ pp in policy accuracy (95\% CI: [3.29, 23.68]; $p=0.0205$); SkillApt activates on 42 rather than 53 states, and both avoid all three observed correctness harms. The comparison supports the outcome-induced policy against one budgeted explicit-boundary implementation; it does not establish superiority over the full BASM system.

\paragraph{D1 discrepancy and D2 correction.}
An unconditional break in the original boundary-budget loop consumed only the first item of each generated field. D1 is preserved in the experimental archive, but is not used as the paper's boundary comparator. D2 serializes complete fields up to the frozen 256-whitespace-token total cap. The same boundary contents and checker configuration are retained; 15/20 full-budget Skills reach the cap, with mean/median/maximum consumption 250.2/256/256 tokens. A boundary-generation failure for \texttt{logicbench\_003} invokes the frozen fallback; excluding that Skill yields D2 policy accuracy $90/144$ (62.5\%) and selected-arm success $129/152$ (84.9\%). At the primary endpoint, policy-label discordance is 44 SkillApt-only correct, 24 D2-only correct, 73 both correct, and eleven both wrong; task-success discordance is ten SkillApt-only correct, one D2-only correct, 136 both correct, and thirteen both wrong. The boundary checker consumes 387,373 recorded tokens across 160 full-budget checker calls; selected-arm task tokens do not include this controller overhead.

\paragraph{WITH-only confusion.}
Replacing paired targets with WITH-only success leaves task success unchanged at $146/160$ (difference zero; 95\% CI: [$-3.13$, $+3.13$] pp) but lowers policy accuracy from $117/152$ to $26/152$ (SkillApt minus WITH-only $+59.9$ pp; 95\% CI: [50.00, 69.08]; $p=4.40\times10^{-20}$), with activation on $159/160$ states and 1,815.2 versus 1,188.5 mean selected tokens. Among 240 evidence states, 204 are jointly correct, 23 improve correctness with the Skill, seven worsen it, and six are jointly incorrect. Of the 204 jointly correct states, 168 are cost-negative, 22 cost-positive, and fourteen neutral. A positive WITH-only label therefore conflates several marginal-utility categories. On 152 non-neutral held-out states, the policy-accuracy comparison has 101 paired-only correct decisions, ten WITH-only-only correct, sixteen both correct, and 25 both wrong. On task success, the policies each win three discordant states, with 143 both correct and eleven both wrong. The requested WITH-only-plus-Cost variant was not run: the frozen cost criterion is relational and requires WITHOUT-arm costs, and introducing a new absolute threshold would change the protocol.

\paragraph{State conditioning and evidence capacity.}
Skill-Global-All uses the unweighted mean target of all evidence of the same Skill and reaches $145/160$ (90.6\%) success, $97/152$ (63.8\%) policy accuracy, $64/160$ (40.0\%) activation, and 1,321.8 selected tokens. SkillApt minus Global-All is $+0.6$ pp in task success (95\% CI $[0,+1.88]$, $p=1$) and $+13.2$ pp in policy accuracy (95\% CI $[+7.90,+19.08]$, $p=1.10\times10^{-5}$). The 22 policy disagreements all switch Global-All \load\ to SkillApt \abstain: 20 cost-negative, one correctness-harmful, and one cost-positive. Global-5 takes only the first five items and reaches $143/160$ (89.4\%) success, $115/152$ (75.7\%) policy accuracy, $40/160$ (25.0\%) activation, and 1,145.1 selected tokens; SkillApt minus Global-5 is $+1.9$ pp in task success (95\% CI $[0,+4.38]$; $p=0.25$) and $+1.3$ pp in policy accuracy (95\% CI $[-5.92,+8.55]$; $p=0.86$), neither significant. Because the global controls also differ in effective evidence capacity, these results support a more selective operating point but do not isolate state conditioning as uniquely necessary.

\subsection{Learned-Applicability Baseline Details}
\label{app:lr}
The logistic-regression baseline (Section~\ref{sec:lr}, Table~\ref{tab:ablation}) is fully offline: no model call, no new execution arm, no modified frozen outcome, and no held-out tuning. Its features $x=[\phi(z),\psi(S_i),\cos]\in\mathbb R^{513}$ reuse SkillApt's deterministic 256-d hashing ($\psi$ hashes the corpus name+description+content with the same rule) and are standardized on the training partition. Labels are the frozen \texttt{utility\_class} binary policy target. Of the 240 evidence states, 45 are \load, 180 \abstain, and 15 neutral-excluded, giving 225 trainable states across 20 Skills with a class ratio of about 1:4.

Selection uses \texttt{GroupKFold}(5) grouped by \texttt{skill\_id} on the 225 non-neutral evidence states, over $C\in\{0.1,1,10\}$ and $\text{class\_weight}\in\{\text{None},\text{balanced}\}$, by mean grouped-CV balanced accuracy; it selects $C=0.1$ with no class weight (tie-break: stronger regularization, then unweighted), and the \texttt{balanced} weight did not improve balanced accuracy. The grouped-CV balanced accuracy at the chosen configuration is $0.614\pm0.056$; a \texttt{GroupShuffleSplit} (test\_size $=0.2$) check with seeds $\{0,1,2,3,4\}$ gives validation balanced accuracy $0.572\pm0.076$ (range $0.505$--$0.719$), only slightly above the $0.5$ chance level, indicating that a reliable global linear boundary is hard to learn from 225 states in 513 dimensions with a 1:4 imbalance. The final model retrains on all 225 non-neutral evidence states and is evaluated once on the 160 held-out states: task success $134/160$ (83.8\%), activation $32/160$ (20.0\%), policy accuracy $115/152$ (75.7\%), mean selected tokens 1,018.5. It loads on only 9 of the 25 utility-positive held-out states (versus 15 for SkillApt). Limitations: one frozen model, two domains, and 225 training states; a global linear classifier versus SkillApt's same-Skill neighbors is a modeling-family difference, not only a capacity difference; and the features are lexical hashing rather than a learned encoder. The result therefore shows only that this specific lightweight global linear baseline does not surpass SkillApt, without a stronger generalization claim.

\subsection{Sequential RSI Replay Details}
\label{app:sequential}
\paragraph{Protocol.}The stream consists of the 240 Phase A evidence states (20 Skills $\times$ 12), ordered by the frozen manifest field \texttt{frozen\_order\_within\_skill}; held-out states are not used. At round $r$, each Skill's current state is scored by the unchanged SkillApt estimator (Eq.~\ref{eq:score}, $\tau=0.5$) using only that Skill's states with order $<r$; the selected arm's frozen outcome is recorded; only then are the state's paired outcomes and target appended to the evidence store. Each state is predicted exactly once. At $r=1$ the evidence is empty, the score equals the prior $0.5$, and the policy abstains. The following checks are asserted in code: every prediction uses exactly $r-1$ earlier states; no held-out state enters any evidence store; a state's own target never enters its own prediction; frozen outcomes are read-only; and permutations reorder states without reading outcomes. Because forming paired evidence still reads the frozen counterfactual arm after the decision, this is a prequential replay, not production online learning.

\paragraph{References and overall rates.}Cold-start always takes the WITHOUT arm and Always Load the WITH arm. The full-evidence reference scores each state by leave-one-out over the other 11 states of its Skill, including later ones; it is an upper reference, not an oracle or a proven optimum. Over the 240 frozen-order predictions, task success and mean selected tokens are: Cold-start $211/240$ (87.9\%), 866.5; Always Load $227/240$ (94.6\%), 1,806.5; sequential SkillApt $221/240$ (92.1\%) with activation $56/240$ (23.3\%), 1,140.5; full-evidence reference $223/240$ (92.9\%) with activation $66/240$ (27.5\%), 1,209.2. Sequential SkillApt rises from $71/80$ (88.8\%) success and 15.0\% activation in rounds 1--4 to $76/80$ (95.0\%) and 27.5\% in rounds 9--12, but round-matched Cold-start also rises from $69/80$ to $71/80$, which is why the main text reports the round-matched gain rather than raw early--late success. The per-round gains $\Delta_r$ (pp) are $0,0,5,5,0,0,15,0,5,15,5,0$ for $r=1,\ldots,12$, giving early and late means of $+2.5$ ($2/80$) and $+6.25$ ($5/80$). The gain over Always Load is $\le 0$ in every round: the evidence stream has a higher WITHOUT-correct base rate than the held-out set, so here SkillApt's value is selectivity at 36.9\% fewer selected tokens than Always Load rather than higher raw success.

\begin{figure*}[t]
\centering\includegraphics[width=\textwidth]{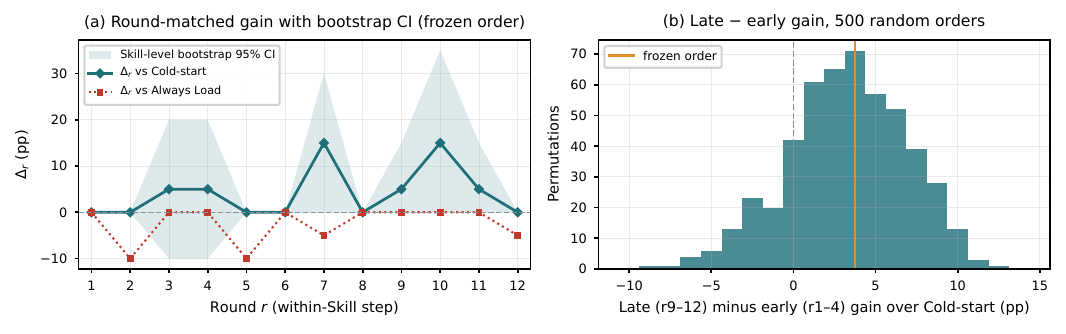}
\caption{Uncertainty in the sequential replay. (a) Round-matched gain over Cold-start in the frozen order with a skill-level bootstrap 95\% interval (2,000 resamples of the 20 Skills), and the gain over Always Load; with one state per Skill per round, $\Delta_r$ moves in 5\,pp steps and the per-round interval is wide. (b) Distribution of the late-minus-early gain over 500 within-Skill random orderings; the vertical line marks the frozen order.}\label{fig:sequential-ci}
\end{figure*}

\paragraph{Order robustness.}Each Skill's 12 states are permuted uniformly at random, independently of outcomes (500 permutations, fixed seed 20260930), and every permutation is replayed with the same test-then-update protocol (Table~\ref{tab:sequential-perm}, Fig.~\ref{fig:sequential-ci}b). Cumulative gain over round-matched Cold-start is positive in all 500 permutations (minimum $+1.7$\,pp). Late gain strictly exceeds early gain in 390 of 500 permutations (78\%), with 42 ties; late raw success exceeds early raw success in 73\%. The early-to-late strengthening is therefore a robust tendency, not a deterministic property, and it flattens or reverses in about a fifth of orderings. The frozen order is not a favorable selection: it lies at the 76th percentile of cumulative success and the 61st percentile of the late-minus-early gain.

\begin{table}[t]
\centering\footnotesize
\setlength{\tabcolsep}{4pt}
\caption{Within-Skill order robustness of the sequential replay (500 outcome-independent permutations; interval = 2.5th--97.5th percentile across permutations). Gains are relative to round-matched Cold-start on identical states.}\label{tab:sequential-perm}
\begin{tabular}{@{}lrr@{}}
\toprule
Quantity & Mean & 95\% interval \\
\midrule
Cumulative success (\%) & 91.8 & [90.4, 92.9] \\
Cumulative activation (\%) & 22.2 & [15.8, 27.5] \\
Cumulative gain (pp) & +3.9 & [+2.5, +5.0] \\
Early gain, rounds 1--4 (pp) & +1.9 & [$-1.3$, +6.3] \\
Late gain, rounds 9--12 (pp) & +5.1 & [0.0, +8.8] \\
Late $-$ early gain (pp) & +3.3 & [$-3.8$, +10.0] \\
\bottomrule
\end{tabular}
\end{table}

\subsection{Cross-Model Diagnostics}
\label{app:cross-stats}
\begin{table*}[t]
\centering\small
\caption{Full matched-replication operating points (complete version of Table~\ref{tab:matched}). Task success is in percent; changes and intervals are in percentage points (pp). Activation and tokens are for the 12-evidence policy.}\label{tab:matched-full}
\begin{tabular}{@{}lrrrrrrr@{}}
\toprule
Model & Zero (\%) & 12 evidence (\%) & $\Delta$ (pp) & 95\% CI (pp) & $p$ & Activation (\%) & Tokens \\
\midrule
GPT-5.6 Terra & 82.5 & 91.3 & $+8.8$ & [$+3.75$, $+15.00$] & 0.0156 & 27.5 & 970.9 \\
Claude Sonnet 4.6 & 65.0 & 70.0 & $+5.0$ & [$-3.75$, $+13.75$] & 0.4240 & 38.8 & 1,222.2 \\
\bottomrule
\end{tabular}
\end{table*}

\begin{table}[t]
\centering\small
\caption{Matched-model activation diagnostics. Policy accuracy excludes model-specific neutral states.}\label{tab:matched-extra}
\begin{tabular}{@{}lrr@{}}\toprule
Metric & Terra & Claude\\\midrule
Useful recall & $7/13$ & $10/18$\\
Cost-negative avoidance & $48/63$ & $39/55$\\
Correctness-harm avoidance & $1/1$ & $0/5$\\
Policy accuracy & $56/77$ & $49/78$\\
Evidence-only task successes & 7 & 9\\
Zero-only task successes & 0 & 5\\
Both correct & 66 & 47\\
Both wrong & 7 & 19\\\bottomrule
\end{tabular}
\end{table}
Terra's 95\% CI is $[+3.75,+15.00]$ pp and Claude's $[-3.75,+13.75]$ pp (Table~\ref{tab:matched-full}). Claude's full-budget policy activates all five of its correctness-harmful states ($0/5$ avoided; Terra $1/1$), its mean selected tokens rise from 759.1 to 1,222.2, and its policy accuracy is $49/78$ at full budget, so the replication establishes neither universal efficiency gains nor population safety rates. These sparse, model-specific denominators prohibit interpreting the percentages as stable population safety rates.

On the matched 80 states, WITH correctness agreement is $66/80$ (82.5\%), WITHOUT agreement $64/80$ (80.0\%), five-class utility agreement $54/80$ (67.5\%), and positive-versus-nonpositive target agreement $63/80$ (78.8\%); 17 states change binary target across models. These findings are consistent with model-conditioned Skill utility, although cost-label differences also reflect measured latency and provider behavior. The target-agreement diagnostic maps utility-positive to \load\ and all other classes to \abstain, including neutral for this particular calculation, whereas policy accuracy excludes neutral rows. The agreement statistics are descriptive and do not demonstrate that a policy trained for one model transfers to the other.

\subsection{Fresh-Cohort Replication Details}
\label{app:fresh}
The cohort in Section~\ref{sec:cross} was built by a coverage audit over historical Terra executions. Thirteen Skills had at least 20 strictly fresh states (task identities disjoint from every prior raw outcome and frozen cohort) and at least one valid same-Skill, same-model historical pair; \texttt{medcalcbench\_018} had no valid pair and was excluded, leaving 12 Skills: \texttt{logicbench\_001}, \texttt{002}, \texttt{005}, \texttt{008}, \texttt{011}, \texttt{013}, \texttt{016}, \texttt{018} and \texttt{medcalcbench\_027}, \texttt{029}, \texttt{047}, \texttt{049}. Historical correctness was not used to select Skills, and none of the 12 appears in Phase A. The 240-state test manifest and the policy configuration were hashed before any request; 480 executor arms and 240 controller decisions were then run on \texttt{gpt-5.6-terra}, all valid (four HTTP 403 records are retained in the raw log and were retried under the same frozen condition). SkillApt's evidence consists of all deduplicated same-Skill Terra pairs in the historical pool (495 pairs over 32 Skills in total): 6, 2, 9, 3, 14, 22, 19, and 11 pairs for the eight LogicBench Skills and 1, 4, 1, and 3 for the four MedCalc-Bench Skills; no fresh outcome was fed back. The zero-shot controller uses the Phase J prompt verbatim. The 240 fresh pairs comprise 26 correctness-positive, 20 correctness-harmful, 7 tied cost-positive, 174 tied cost-negative, and 13 neutral states.

Table~\ref{tab:fresh} lists the policies. SkillApt minus zero-shot is $+2.92$ pp (task-level paired bootstrap 95\% CI $[-1.25,+7.50]$; Skill-cluster $[-3.75,+12.50]$; 18 SkillApt-only versus 11 zero-shot-only correct states, exact McNemar $p=0.265$), and SkillApt minus cold start is $+5.0$ pp (15 versus 3). Per domain, LogicBench gives 136 (SkillApt), 138 (zero-shot), 124 (cold start), and 140 (always load) of 160, with SkillApt loading on 74 states; MedCalc-Bench gives 73, 64, 73, and 63 of 80, with SkillApt loading on none, because one to four historical pairs never moved the score above the prior. A utility-gated variant evaluated in the same run made nearly identical decisions (210/240 correct, 69/240 loads) and is not part of this paper's method. These are descriptive matched results on one executor; they support the direction of the Phase A comparison and the value of abstaining under sparse evidence, not a superiority claim.

\begin{table}[t]
\centering\scriptsize
\setlength{\tabcolsep}{3pt}
\caption{Fresh-cohort replication (12 Skills disjoint from Phase A, 240 fresh states). Harm counts the 20 correctness-harmful states on which a policy chose \load; Recall is useful recall over the 33 utility-positive states. Selected tokens are omitted for Cold Start and Always Load because the processed artifact records only the arms selected by the two controllers.}\label{tab:fresh}
\begin{tabular}{@{}lrrrrr@{}}
\toprule
Policy & Success & Activ. & Harm & Recall & Tokens \\
\midrule
Cold Start & 197/240 & 0/240 & 0/20 & 0/33 & -- \\
Always Load & 203/240 & 240/240 & 20/20 & 33/33 & -- \\
Zero-shot LLM & 202/240 & 229/240 & 19/20 & 31/33 & 1,733.9 \\
SkillApt & 209/240 & 74/240 & 3/20 & 15/33 & 1,200.5 \\
\bottomrule
\end{tabular}
\end{table}

\subsection{Retrieval-Based Confirmatory Study}
\label{sec:retrieval}
The supporting retrieval study compares Never Load, Gold Always, BM25 Top-1, Semantic-only, SkillApt, and SkillApt-Fusion. Its canonical BM25 uses Skill name, description, and content in a 26,262-Skill corpus, with $k_1=1.5$, $b=0.75$, and $K=10$; its earlier evidence split and confirmatory set are separate from Phase A, and the two protocols are not pooled. Canonical BM25 retrieves the target Skill with R@1/R@5/R@10 of 0.239/0.389/0.513 and MRR@10 of 0.301 on 113 frozen targets. Two infrastructure-censored states leave 111 complete policy evaluations. SkillApt and BM25 Top-1 both achieve observed accuracy 83.8\%, while activation falls from 100.0\% to 31.5\% and mean selected tokens from 4,782 to 1,231, approximately 74.3\% lower.

The paired accuracy difference is zero, with 95\% CI [$-6.31$, $+6.31$] pp and exact McNemar $p=1.0$. This is the same \emph{observed} accuracy, not statistical equivalence. Never Load obtains 79.3\%, Gold Always 88.3\%, Semantic-only 84.7\%, and SkillApt-Fusion 86.5\%. Semantic-only and fusion both activate throughout, so they occupy substantially higher-exposure operating points. This supporting result demonstrates potential retrieval-stage efficiency under one frozen setting; it does not generalize the primary cold-start comparison's cost direction.

\paragraph{Retrieval provenance and failed stronger condition.}
The canonical retrieval representation concatenates name, description, and content. A subsequent metadata-only BM25 experiment used name and description; its numbers do not replace the canonical R@1/R@5/R@10 above. In the preregistered stronger-retriever control, canonical R@10 was 0.5133 and the fixed reranker reached 0.2124. The stronger condition was therefore not established and the planned downstream comparison stopped. This is a failed prerequisite, not evidence of robustness to stronger retrieval.

\paragraph{Pre-outcome calibration-extension stop.}
A preregistered richer paired-benefit calibration extension (Phase M) would have resolved correctness ties through blinded artifact-quality comparisons. Its sole compliant frozen-configuration validation used 20 fixed development pairs and 40 logical comparisons: 36/40 responses were parseable, 16 pairs were valid bidirectionally, and 12/16 were swap-consistent (75\%, below the prespecified 90\% threshold). The extension stopped with \texttt{QUALITY\_JUDGE\_UNSTABLE}, before any q0 prediction or new executor outcome (0/184 planned arms). This is a prerequisite failure, not an efficacy result for that calibration variant. The earlier diagnostic sent 16 rather than the frozen 64 maximum output tokens; it is retained only as a noncompliant diagnostic, not a second valid judge test.

\subsection{Token-Only Cost Sensitivity}
\label{app:token-only}
Because tool calls are zero in the direct-generation benchmark and latency includes provider noise, we recompute the correctness-tie label with tokens alone, $g_i^{\mathrm{tok}}=(C^{(0)}_{\mathrm{tok}}-C^{(1)}_{\mathrm{tok}})/\max(1,C^{(0)}_{\mathrm{tok}})$, keeping the $\pm0.05$ thresholds of Eq.~\ref{eq:cost}. Correctness labels, the estimator, task success, and all policy decisions are unchanged, and no model is called; the original multi-resource labels are first reproduced exactly from the raw runs. On the 199 Phase B triplets (benefit/harm/cost-positive/cost-negative/neutral), the Hard Candidate breakdown moves from 6/12/5/169/7 to 6/12/0/181/0 and the Gold Skill breakdown from 13/9/14/142/21 to 13/9/1/173/3. All 181 Hard Candidate correctness ties increase tokens (169 were cost-negative under the multi-resource rule), as do 176 of the 177 Gold ties. On the 160 Phase A held-out states, token-only relabeling gives 136 tied cost-negative states; all 124 original ones remain cost-negative. Cost-negative avoidance is $109/136$ for SkillApt (originally $99/124$), $2/136$ for the zero-shot LLM controller ($2/124$), and $1/136$ for WITH-only ($1/124$).

The direction therefore agrees with, and slightly strengthens, the multi-resource taxonomy: correctness-tied exposure mostly adds cost, SkillApt avoids most of it, and the always-loading controllers do not. The category that is not robust is cost-positive: none of the 14 Gold, 5 Hard Candidate, or 29 Phase A cost-positive ties remains cost-positive under tokens alone, so their advantage came from latency. Because the estimator maps every tie to $1/2$, SkillApt's scores and decisions are unaffected; only diagnostics that count cost-positive ties as utility-positive change. SkillApt's 15 useful loads are all correctness-positive (15 of 18; none of the 7 held-out cost-positive ties), whereas the logistic-regression baseline's 9 are 4 correctness-positive and 5 cost-positive. Under token-only labels, useful recall is $15/18$ for SkillApt versus $4/18$ for logistic regression, and policy accuracy is $127/157$ versus $118/157$, so the comparison in Section~\ref{sec:boundary} does not depend on latency-driven labels.

\section{Real-World Software-Agent Lifecycle Evidence}
\label{app:realworld}
An anonymized real-world software-agent execution environment provides the software-engineering data used for lifecycle evidence and small controlled fixtures (Section~\ref{sec:realworld}). This part is supplementary external-validity evidence only and is not training data for the SkillApt estimator. A deterministic miner extracted 76 episodes from two applications, with rules and gold labels developed on the same samples. We therefore do not treat miner accuracy as out-of-sample generalization. The controlled component uses nine fixtures, two conditions, and three repetitions: 54 designed runs. Both arms share the fixture, pre-state, validator, tool availability, and prompt baseline.
\begin{table}[t]
\centering\footnotesize
\setlength{\tabcolsep}{3.5pt}
\caption{Real-world software-agent lifecycle evidence: matched WITHOUT/WITH fixture families (success as WITHOUT $\rightarrow$ WITH; token change compares medians). Evidence base: 76 mined historical episodes and 54 controlled microbenchmark runs. These are small controlled fixtures, not production-wide longitudinal validation, and they are not training data for the SkillApt estimator. \textsuperscript{a}Treatment-attributable runs (12:11) after excluding one WITH run that did not load the target.}\label{tab:lifecycle}
\begin{tabular}{@{}lrrr@{}}
\toprule
Software-agent case & Runs & Success & Median \\
 & & & tokens \\
\midrule
Dataset-contract repair & 23\textsuperscript{a} & 100\% $\rightarrow$ 100\% & $-18\%$ \\
Global-import cleanup & 30 & 80\% $\rightarrow$ 100\% & $-31\%$ \\
\bottomrule
\end{tabular}
\end{table}

For dataset-contract repair, 23 treatment-attributable runs retain 100\% success with about 18\% lower median tokens; for global-import cleanup, 30 runs improve success from 80\% to 100\% with about 31\% lower median tokens (Table~\ref{tab:lifecycle}). Of 27 WITH runs, 26 demonstrably load the target; excluding the single non-loading dataset-contract run leaves 23 attributable runs for that family. Dataset-contract token medians are 203,892 and 166,492; global-import medians are 1,051,919 and 729,809. Per-repetition spread can reach 133\% of a fixture/arm median. These measurements should therefore be interpreted as local directional evidence, not a precise general efficiency effect. The global-import success difference comes from two of five fixtures whose unassisted arms pass only two and one of three repetitions respectively, while assisted arms pass all three.

Separately, seven episodes across four applications concern a validator family that no longer triggers. No paired effect estimate is assigned to that invalid path. This motivates environment-validity checks rather than interpreting historical frequency as current usefulness.

The larger temporal-replay preparation identified sufficient recurring Skill use and selected 48 early plus 48 later candidate states for six Skills (192 planned arms). The sample lacked exact pre-session checkpoints, task-specific evaluators, a verified current Agent Team harness, complete Skill hashes, and frozen tool-service controls. It stopped before any outcome API call. These are feasibility observations only; no temporal task-success result, production-level longitudinal RSI, trajectory-based policy learning, or autonomous improvement result is claimed.

\end{document}